\UseRawInputEncoding
\documentclass[prd,aps,letterpaper,floatfix,superscriptaddress,preprint,11pt,nofootinbib]{revtex4-1}
\usepackage{epsfig,epsf}
\usepackage{booktabs}
\usepackage{mathrsfs}
\usepackage{graphicx}
\usepackage{subcaption}
\usepackage{caption}
\usepackage{latexsym}
\usepackage{amsmath}
\usepackage{amssymb}
\usepackage{textcomp}
\usepackage{amsbsy}
\usepackage{tabularx}
\usepackage{slashed}
\usepackage{array}
\usepackage[dvipsnames]{xcolor}
\usepackage{bm}
\usepackage{amsmath,amsfonts,amssymb}
\usepackage{hyperref}
\DeclareGraphicsExtensions{.png,.eps,.pdf}

\usepackage{color}
\usepackage[normalem]{ulem}

\definecolor{darkgreen}{cmyk}{1,0,1,0.4}

\long\def\/*#1*/{}

\begin{document}
\title{Explorations of pseudo-Dirac dark matter having keV splittings and interacting via  transition electric and magnetic dipole moments}
\author{Shiuli Chatterjee}
\email{Shiuli.Chatterjee@ncbj.gov.pl}
\affiliation{National Centre for Nuclear Research, Pasteura 7, 02-093 Warsaw, Poland}
\author{Ranjan Laha}
\email{ranjanlaha@iisc.ac.in}
\affiliation{Centre for High Energy Physics, Indian Institute of Science, Bangalore 560012, India}
\date{\today}
\begin{abstract}
We study a minimal model of pseudo-Dirac dark matter, interacting through transition electric and magnetic dipole moments. 
Motivated by the fact that  xenon experiments can detect electrons down to $\sim$\,keV recoil energies, we consider $\mathcal{O}$(keV) splittings between the mass eigenstates. 
We study the production of this dark matter candidate via the freeze-in mechanism. We discuss the direct detection signatures of the model arising from the down-scattering of the heavier state, that are produced in Solar upscattering, finding observable signatures at the current and near-future  xenon based direct detection experiments. 
We also study complementary constraints on the model from fixed target experiments, lepton colliders, supernovae cooling and cosmology.
We show that the latest XENONnT results rule out parts of the parameter space for this well motivated and minimal dark matter candidate. Next generation  xenon experiments can either discover or further constrain how strongly inelastic dark matter can interact via the dipole moment operators.
\end{abstract}
\maketitle

\section{Introduction}
The identity of dark matter (DM) is one of the most important questions in science. Multiple broad approaches are pursued in order to address this important question\,\cite{Green:2021jrr,Cooley:2021rws,Slatyer:2021qgc}.
One of the frontiers of DM physics today can be understood to lie along the reach of current detection methodologies looking to identify the particle nature of DM and/ or its interactions with Standard Model (SM) particles, pushed farther along by the projections of near future detectors.  From this perspective, 
hints of anomalies in data as well as
new detectors that achieve improved efficiency and background control make for exciting progress and spur new investigations into the physics of DM. One such result came from the XENON1T experiment\,\cite{XENON:2020rca} that reported a $3\sigma$ excess in the electron recoil spectrum between $2-3$\,keV and attracted wide attention from the perspective of DM interpretation\,\cite{An:2020bxd,Baryakhtar:2020rwy,Keung:2020uew,Takahashi:2020bpq,Choudhury:2020xui,Harigaya:2020ckz,Cao:2020bwd,Fornal:2020npv,Ko:2020gdg,Su:2020zny,Alonso-Alvarez:2020cdv,Lee:2020wmh,Bramante:2020zos,Paz:2020pbc,Nakayama:2020ikz,Bell:2020bes,Chen:2020gcl,Chao:2020yro, Smirnov:2020zwf, Kannike:2020agf, Boehm:2020ltd, Du:2020ybt, Choi:2020udy, Buch:2020mrg, Dey:2020sai, Jho:2020sku, Bloch:2020uzh, Lindner:2020kko, Budnik:2020nwz, Zu:2020idx, Gao:2020wer, DeRocco:2020xdt, Dent:2020jhf, McKeen:2020vpf, DelleRose:2020pbh, Alhazmi:2020fju, An:2020tcg, Croon:2020ehi, Chigusa:2020bgq, Okada:2020evk, Davighi:2020vap, Choi:2020kch, Baek:2020owl, Davoudiasl:2020ypv, Chiang:2020hgb, He:2020wjs, Long:2020uyf, Croon:2020oga, Arcadi:2020zni, Ema:2020fit, Khan:2020pso, Farzan:2020llg, Zu:2020bsx, Lasenby:2020goo, Chakraborty:2020vec, Guo:2020oum, Aboubrahim:2020iwb, Buttazzo:2020vfs, Choi:2020ysq, He:2020sat, Jia:2020omh, Xu:2020qsy, Dutta:2021wbn, Bell:2021zkr, Dutta:2021nsy, Baek:2021yos,PandaX-II:2020udv,Dror:2019dib,Dror:2019onn,Dror:2020czw,Borah:2020jzi,Borah:2021yek,Emken:2021vmf,Borah:2021jzu} or for background considerations\,\cite{Robinson:2020gfu, Bhattacherjee:2020qmv, Szydagis:2020isq}. It also reported a background rate of $76\pm 2$\,events$/(\textrm{tonne}\times\textrm{year}\times\textrm{keV})$ for electron recoil energies between $1-30$\,keV. Subsequently, this background was reduced to $15.8\pm 1.3$ \,events$/(\textrm{tonne}\times\textrm{year}\times\textrm{keV})$ by the XENONnT experiment \cite{XENONCollaboration:2022kmb}, while observing no excess, with an exposure of 1.16 tonne$\times$years.

The scattering of typical WIMP-like DM, with velocity distributions following the Standard Halo Model, against electrons lead to recoil energies of $\mu_{e,DM}v_{DM}^2\sim \mathcal{O}$(eV)\footnote{See ref.\,\cite{Bloch:2020uzh} for a discussion on the recoil energies in DM scattering against electrons inside a nucleus. For comparison, a DM particle with a form factor of $F_{DM}(q)=1$ is shown to be unable to explain the excess without being in conflict with XENON1T S2-only analysis. While DM with form factors $F_{DM}(q)\propto q,q^2$ mediated by heavy particles are shown to be give recoil energies of $\mathcal{O}$(keV).}, where $\mu_{e,DM}$ is the electron-DM reduced mass and $v_{DM}$ is the incoming DM velocity. The $\mathcal{O}$(keV) recoil energies that XENON1T probes currently can broadly arise from various DM models like boosted DM\,\cite{Fornal:2020npv,Su:2020zny,Cao:2020bwd,Ko:2020gdg,Chen:2020gcl} that acquire higher velocities than the typical DM halo velocities, absorption of $\mathcal{O}$(keV) particles\,\cite{Takahashi:2020bpq}, or through down-scattering of inelastic DM with two nearly degenerate DM states with mass splittings of $\mathcal{O}$(keV)\,\cite{Baryakhtar:2020rwy,Keung:2020uew,Harigaya:2020ckz,Lee:2020wmh,An:2020tcg,Chao:2020yro}. 
The coincidence of similarity in the recoil energies probed by XENONnT and the temperature at the Sun's core ($\sim$ 1.1 keV) motivates the study of inelastic DM which is upscattered in the Sun. 
Specifically, if the heavier particle is not cosmologically stable, and the lighter particle constitutes the DM, they can  be excited to the heavier particle by upscattering against Solar electrons. It can subsequently down-scatter via electron scatterings at XENONnT, depositing an energy approximately equal to the mass splitting.

With this in mind, we study the most minimal model of pseudo-Dirac DM with two Majorana fermions. The lowest dimension interactions with SM allowed, in the absence of any additional dark sector particles, are the transition electric and magnetic dipole moments.
Dipolar DM \cite{Masso:2009mu,Blanchet:2009zu} with scattering processes mediated by the photon, gives enhanced scattering rates in the small velocity limit applicable in terrestrial direct detection\,(DD) experiments\,\cite{Hambye:2018dpi}.
We therefore study:
\begin{itemize}
    \item The freeze-in\,(FI) production\,\cite{McDonald:2001vt,Hall:2009bx} of pseudo-Dirac DM with mass splittings of $\mathcal{O}$(keV).
    \item The DD of this DM via up-scattering in the Sun followed by down-scattering in electron recoil events at the XENONnT experiment (as well as projections for future runs of  XENONnT and DARWIN). 
    We study the DD rates for generic parts of the parameter space without the relic density constraint from FI production. 
    \item Complementary bounds on generic parts of the parameter space. These include constraints from fixed target experiments, lepton colliders, SN1987A and $N_{\rm eff}$ constraints.
\end{itemize}
Inelastic DM with interaction via transition dipolar moments have previously been studied with focus on different parameter spaces\,\cite{Weiner:2012cb,Patra:2011aa,Masso:2009mu}. Refs.\,\cite{CarrilloGonzalez:2021lxm,Filimonova:2022pkj,Herrera:2023fpq} have also studied pseudo-Dirac type DM in a dark photon model. Dipolar DM motivated by lepton sector minimal flavor violation was recently studied in\,\cite{DAmbrosio:2021wpd}. 

In section\,\ref{sec:FI}, we discuss our model and the production of the DM via FI mechanism. In section\,\ref{sec:DD}, we discuss the signatures at DD experiments from electron and nuclear scatterings. In section\,\ref{sec:CB}, we discuss complementary bounds on the model from various existing experiments and observations.  Finally, we present the results in section\,\ref{sec_res} and conclude in section\,\ref{sec:conclusions}.
\section{Model Framework and Freeze-in Production}\label{sec:FI}
We consider a dark sector consisting of two Majorana fermions $\chi_1$ and $\chi_2$ that form a pseudo-Dirac state with mass splitting $\delta\equiv m_{\chi_2}-m_{\chi_1}\sim \mathcal{O}(\textrm{keV})$. The lowest dimension interaction operators allowed in the absence of any additional new particles are through the DM dipole moments where the interaction is mediated by photons. Since the only dipolar type interactions that are allowed for Majorana fermions are of transition type, we get the following Lagrangian for DM interaction\,\cite{Masso:2009mu}
\begin{equation}\label{lagrangian}
    \mathcal{L}_{DM}\supset \underbrace{\frac{\mu_\chi}{2}\bar{\chi}_1\sigma_{\mu\nu}\chi_2\, F^{\mu\nu}}_{MDM} +\underbrace{i\frac{d_\chi}{2}\bar{\chi}_1\sigma_{\mu\nu}\gamma_5\chi_2\,F^{\mu\nu}}_{EDM} ,
\end{equation}
where $\mu_\chi$ and $d_\chi$ are the transition magnetic dipole moment (MDM) and electric dipole moment (EDM), respectively. 

\begin{figure}
    \centering
    \begin{subfigure}[b]{0.40\textwidth}
    \includegraphics[width=\textwidth]{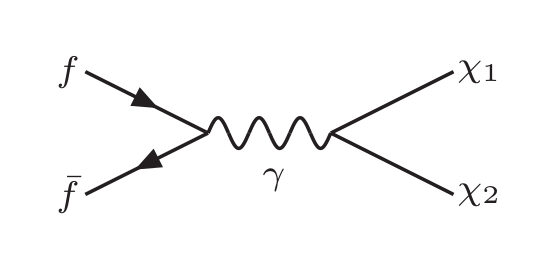}
    \caption{}
    \label{fig:relic_A}
    \end{subfigure}
    \begin{subfigure}[b]{0.40\textwidth}
    \includegraphics[width=\textwidth]{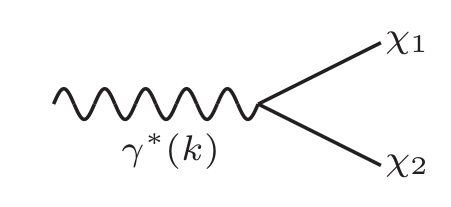}
    \caption{}
        \label{fig:relic_H}
    \end{subfigure}
\caption{Two example processes for FI production of dipolar DM: (a) $2\rightarrow 2$ annihilation production from SM fermions (we do not show here the diagram for production from  $W^+W^-$) and (b) production from plasmon decay.}
\label{fig:ch4_relic}
\end{figure}

We consider the production of DM by the FI mechanism which is operative when the DM coupling to SM is small enough for the DM to have never entered thermal equilibrium with the SM bath. With an assumption of negligible DM number density at the earliest epoch, it is produced via annihilation and decay of SM particles; some relevant Feynman diagrams for dipolar DM are shown in fig.\,\ref{fig:ch4_relic}. 
As the temperature of SM bath falls, the DM production ceases as follows. For $m_{DM}>m_{SM}$, as the bath temperature falls below the DM mass, the DM production becomes kinematically suppressed. While for $m_{SM}>m_{DM}$, as the bath temperature falls below the SM particle mass, it freezes out and its annihilation rate is suppressed. Thus the DM number density per comoving volume becomes a constant, and the DM is said to have frozen into the current relic abundance.\\  

We solve the Boltzmann equation to calculate the DM relic density. For convenience, we define the number per co-moving volume as Yield, $Y\equiv n/s$, with $n$ representing the number density and $s$ being the entropy density. The Boltzmann equation then gives\,\cite{McDonald:2001vt,Hall:2009bx,Dutra:2019gqz}
\begin{equation}\label{eq:yield}
\frac{dY}{dT}=-\left(\frac{45}{\pi}\right)^{3/2}\frac{M_{ Pl}}{4\pi^2g_{\star}^s\sqrt{g_{\star}}}\left(1+\frac{1}{3}\frac{d\,ln\,g_{\star}}{d\,ln\,T}\right)\frac{R(T)}{T^6},
\end{equation}
where T is the temperature of the SM bath, $M_{Pl}$ is the Planck mass, $g^s_*$ and $g_*$ are the temperature dependent relativistic degrees of freedom contributing to entropy and energy densities, respectively. $R(T)$ is the rate for DM production and is a sum of rates  of production from $2\rightarrow 2$ annihilation, $R_{2\rightarrow 2}$, and from the decay of plasmon, $R_{\gamma^*}$,
\begin{equation}
R(T)=R_{2\rightarrow 2}(T)+R_{\gamma^*}(T).
\end{equation}
We discuss these two modes of production in the following subsections. 

The total DM relic density is 
\begin{equation}
\Omega\,h^2=\frac{Y(T_0)\,s(T_0)m_{\chi_1}h^2}{\rho_{crit}}=Ym_{\chi_1}h^2\frac{2.89\times 10^9\,\text{m}^{-3}}{10.5\,h^2\,\text{GeVm}^{-3}}=Ym_{\chi_1}\frac{2.89\times 10^9\,\text{m}^{-3}}{10.5\,\text{GeVm}^{-3}},
\end{equation}
where we have assumed only the lighter particle $\chi_1$ with mass $m_{\chi_1}$ to be stable on cosmological timescales. Here, $\rho_{crit}$ is the critical density and $T_0$ is the temperature of current epoch, with the values for the physical constants taken from ref.\,\cite{Belanger:2018ccd}. 
\subsection{Annihilation production}\label{sec_relic}
The rate of production of particles $\chi_1,\chi_2$ from the annihilation of SM fermions, as shown in fig.\,\ref{fig:ch4_relic} (a), is \cite{Hall:2009bx,Bernal:2017kxu,Dutra:2019gqz}
\begin{eqnarray}
R_{2\rightarrow 2}(T)&=&\sum_f N_c^f\frac{T}{(2\pi)^6 2^4}\int_{s_{min}}^{\infty} ds \sqrt{s-4m_f^2} \frac{|\vec{p}_3^{\,\,0}|}{\sqrt{m_{\chi_1}^2+|\vec{p}_3^{\,\,0}|^2}+\sqrt{m_{\chi_2}^2+|\vec{p}_3^{\,\,0}|^2}} \nonumber \\
&&\times \,K_1\left(\frac{\sqrt{s}}{T}\right)   \int d\Omega_3^{\star}\,|\mathcal{M}|^2,
\end{eqnarray}
where $N_c^f$ are the color degrees of freedom of SM fermion $f$ with mass $m_f$. The total rate is a sum over the SM fermions $f$. $K_1$ is the modified Bessel function of the second kind,  $s_{min}=\textrm{Max}\left[(m_{\chi_1}+m_{\chi_2})^2,4m_f^2\right]$, and 
the 3-momentum of $\chi_1$ in the center-of-mass\,(CM) frame is
$$|\vec{p}_3^{\,\,0}|=\sqrt{\left(\frac{\sqrt{s}}{2}+\frac{m_{\chi_1}^2-m_{\chi_2}^2}{2\sqrt{s}}\right)^2-m_{\chi_1}^2}.$$
The squared-amplitude $|\mathcal{M}|^2$ is {\it summed} over initial and final spins\,\cite{Masso:2009mu}
\begin{equation}\label{amp2}
    \int d\Omega_3^{\star}\sum \big|\mathcal{M}\big|_f^2\simeq
    \begin{cases}
    32\pi d_\chi^2e^2 q_f^2 \frac{(s-4m_{\chi_1}^2)(s+2m_f^2)}{3s}-128 \pi\, \delta d_\chi^2 e^2 q_f^2\frac{m_{\chi_1}(s+2m_f^2)}{3s}+\mathcal{O}(\delta^2) ,& \text{ for EDM,}\\
    32\pi \mu_\chi^2e^2 q_f^2 \frac{(s+8m_{\chi_1}^2)(s+2m_f^2)}{3s}+64 \pi\,  \delta\mu_\chi^2 e^2 q_f^2\frac{m_{\chi_1}(s+2m_f^2)}{3s}+\mathcal{O}(\delta^2),              & \text{ for MDM,}
    \end{cases}
\end{equation}
where we give the expressions upto leading order in $\delta\equiv m_{\chi_2}-m_{\chi_1}$. 
In the limit of vanishing SM fermion and DM masses, we find that the production rate scales with temperature as $R(T)\propto T^6$. From eq.\,\eqref{eq:yield}, we see that this leads to UV (UltraViolet) FI with maximum yield produced at the largest temperature of the SM bath, equal to the reheating temperature for models with instantaneous reheating\,\cite{Chen:2017kvz,Elahi:2014fsa,Chowdhury:2018tzw,Giudice:2000ex}. 

Of phenomenological interest is the parameter space with large coupling which can lead to large scattering rate at DD experiments. We note that for $m_{DM}\geq T_{RH}$, there is kinematic suppression in DM production, requiring large couplings to reproduce the observed relic density. For the light DM we consider in this work, $m_{DM}\leq 1\,{\rm GeV}$, we choose low reheating temperatures ($T_{RH}$\,=\,5\,MeV and 10\,MeV) from allowed\footnote{A robust lower bound on reheating temperatures of $T_{RH}>4$\,MeV was derived by combining cosmic microwave background, large scale structure and light element abundances data in ref.\,\cite{Hannestad:2004px}.}  values.
We verify our calculations using the expressions given above, with those obtained from {\tt micrOMEGAs 5.0}\,\cite{Belanger:2018ccd}, for fermionic channels of production.

We discuss the results in section\,\ref{sec_res}.
\subsection{Plasmon decay}\label{plasmon}
The SM bath consists of charged particles that couple to the photon. An electromagnetic wave propagating through this bath behaves qualitatively differently from that in vacuum. 
Coherent vibrations of the electromagnetic field and the density of charged particles results in a spin-1 particle with 1 longitudinal and 2 transverse polarizations, propagating at a speed less than the speed of light in vacuum. Its dispersion relation depends on the properties of the SM plasma and is therefore known as the `plasmon'.
We follow the discussion in refs.\,\cite{Braaten:1993jw,Raffelt:1996wa} for the plasmon properties and decay rates. 

The significance of plasmon decay leading to production of DM in FI scenarios was noted in ref.\,\cite{Dvorkin:2019zdi} (see also \cite{Chu:2019rok,Chang:2019xva,Hambye:2019dwd}). The massive plasmon can decay to DM in cases where the DM couples to the photon. This is true in the dipolar interaction case we study here.
For DM produced via FI mechanism, the final abundance is accumulated over time and summed over the various production channels of SM particles annihilating and decaying to DM. Therefore, the production from plasmon decay can play a significant role in FI produced DM (as opposed to freeze-out production).

To calculate the DM abundance resulting from plasmon decay, we must begin with the modified dispersion relations which depend on the temperature $T$ and the net density of the charged particles in the SM plasma. 
Since ref.\,\cite{Braaten:1993jw} was addressing the energy loss from the plasma in a supernova with typical energies of $\mathcal{O}(10)$\,MeV, only electrons and positrons were considered as charged particles of relevance that modify the dispersion relations. But for the case of a UV FI with large enough $T_{RH}$, all charged particles with masses much smaller than the reheating temperature would add to the plasmon effect (see appendix\,\ref{app_plasmon} for further details).
The rate of DM production from plasmon decay is
\begin{figure}[b!]
    \centering
    \includegraphics[scale=0.65]{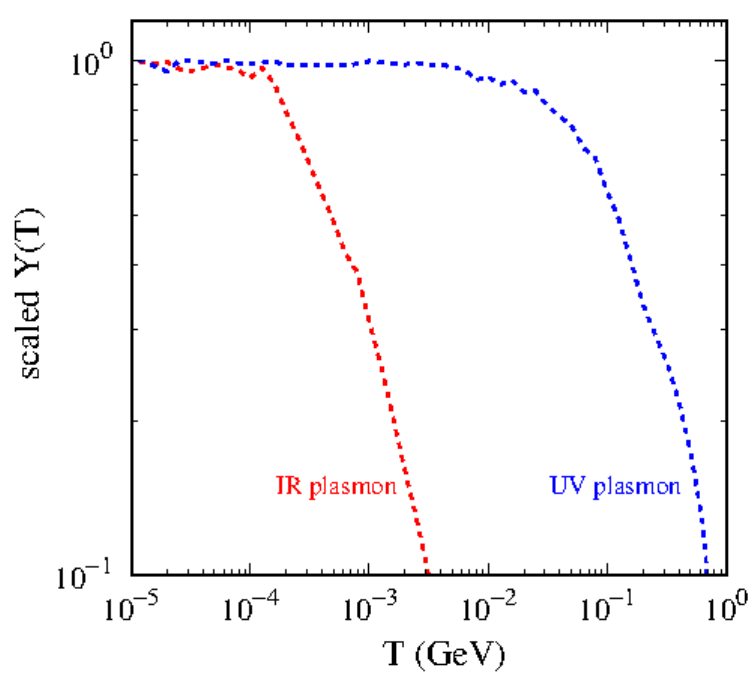}
    \caption{Production of DM from plasmon decay, for IR and UV ($T_{RH}=1$\,GeV) FI models. We consider $m_{DM}=1\,$keV here.}
    \label{fig:ch4_IR_UV_plasmon}
\end{figure}
\begin{eqnarray}
R_{\gamma^*}(T)&=&\sum_{\textrm{pol}=T,L}g_{\textrm{pol}}\int \frac{d^3k}{(2\pi)^3}f(\omega_{\textrm{pol}}) \Gamma_{\textrm{pol}},
\end{eqnarray}
where we assume that only the lighter particle $\chi_1$ survives.
The transverse and longitudinal polarizations are $g_T=2$ and $g_L=1$, respectively. The distribution for the plasmon is given by $f(\omega_{\rm pol})=1/({\rm exp}\frac{\omega_{\rm pol}}{T}-1)$.

The decay width of a plasmon $\Gamma_{\textrm{pol}}$ with four momentum $k=(\omega,\vec{k})$ in the medium frame, with a definite polarization `pol' is \cite{Raffelt:1996wa}
\begin{equation}
    \Gamma_{\text{pol}=T,L}=\int \frac{d^3p_{\chi_2}}{(2\pi)^3\,(2E_{\chi_2})}\frac{d^3p_{\chi_1}}{(2\pi)^3\,(2E_{\chi_1})}(2\pi)^4\delta^4\left(k-p_{\chi_1}-p_{\chi_2}\right)\frac{1}{2\omega_{\textrm{pol}=T,L}}\sum_{\text{spins}} |\mathcal{M}|_{\gamma^*_{\textrm{pol}}\rightarrow \chi_1,\chi_2}^2,
\end{equation}
with $|\mathcal{M}|_{\gamma^*_{\text{pol}}\rightarrow\chi_1\chi_2}^2$ being the squared amplitude for plasmon decay to DM, expressions for which are given in eq.\,\eqref{plasmon_modm2}. For dipolar DM we find that the rate of DM production from plasmon decay is also maximised at the largest temperatures of the SM bath. 
We call this a UV plasmon production. 
We show in fig.\,\ref{fig:ch4_IR_UV_plasmon}, the scaled DM yield as a function of temperature, from plasmon production for dipolar interaction (UV plasmon) and millicharged interaction\,\cite{Dvorkin:2019zdi,Dvorkin:2020xga} (Infrared, IR, plasmon). Here, we have considered a DM of mass 1 keV for both the cases. 
The UV plasmon production can be seen to be maximum at the reheating temperature, taken to be 1 GeV in this figure. As $T_{RH}$ is changed, the IR produced yield is unchanged. While the UV plasmon line would shift towards right (higher temperatures) as $T_{RH}$ is increased .

Although we have only shown the yield for EDM interacting DM in fig.\,\ref{fig:ch4_IR_UV_plasmon} for the UV plasmon, the same is also true for MDM interaction. Note that we show the UV plasmon production for $T_{RH}=1\,{\rm GeV}$ to clearly show the difference in IR vs UV production; the maximum production of DM happens at the largest temperatures even for smaller values of $T_{RH}$.

This is in contrast with IR plasmon production where the 
maximum production takes place at the lowest temperatures so that the plasmon production mechanism begins to dominate for small DM masses, $m_{DM}\lesssim 400$\,keV\,\cite{Dvorkin:2019zdi}. This can be understood by noting that the $2\rightarrow 2$ annihilation process becomes ineffective when electrons freeze out, while the plasmon production continues to be effective. 
While for UV plasmon production (with small $T_{RH}$), the plasmon decay production does not take over the annihilation production as we decrease the DM mass, since both the processes maximize at the largest temperatures of the SM bath. For the small reheating temperatures we consider, mindful of the sub-GeV DM masses that are of interest to the DD discussed in the following, the plasmon production is always subdominant and can be safely ignored. For much larger reheating temperatures $T_{RH}\gtrsim 100$ GeV though, the plasmon production can dominate and must be taken into account.
\section{Direct Detection}\label{sec:DD}
We discuss the DD of inelastic DM with interactions via transition EDM and MDM operators.
As mentioned above, an $\mathcal{O}$(keV) splitting is of interest for production of the excited state in the Sun\,\cite{Baryakhtar:2020rwy}, as well as for detection at electron scattering experiments. Additionally, mediation by the photon leads to an enhanced scattering cross section at low velocities\,\cite{Hambye:2019dwd}. 

\begin{figure}[ht!]
    \centering
    \includegraphics[scale=0.5]{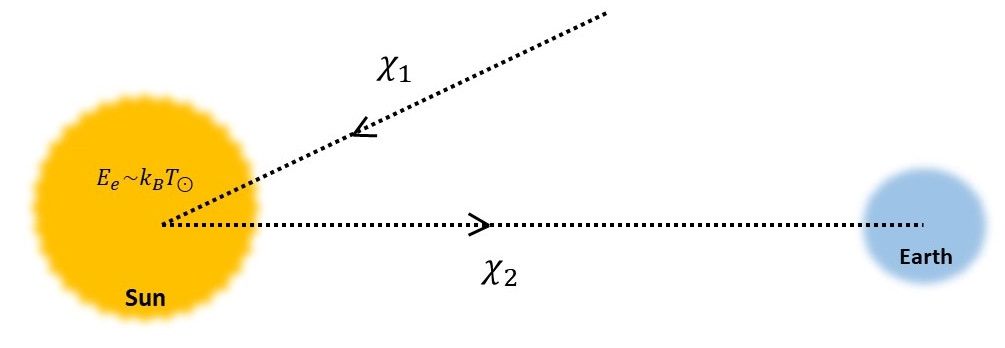}
    \caption{Direct detection of DM up-scattered in the Sun (adapted from \cite{An:2017ojc}).}
    \label{fig:ch4_upscatter}
\end{figure}

For couplings of interest that reproduce the observed relic density, the heavier state $\chi_2$ is not stable on cosmological scales and $\chi_1$ makes up the full DM density. We assume that this is also true for other parts of the parameter space. For the given model, a scattering process can either proceed through the inelastic scattering process of $\chi_1e\rightarrow \chi_2e$ or through a loop-suppressed elastic scattering process, $\chi_1e\rightarrow \chi_1e$. 
The up-scattering process requires larger DM velocities than those allowed in the Galactic halo, while the elastic scattering rate is loop-suppressed. The only avenue\footnote{Cosmic ray upscattering rates will be suppressed since the constituent particles have relativistic speeds and the dipolar scattering rates are inversely proportional to the DM-target relative velocity\,(see eq.\eqref{eq:dsigder}).} for DD is then from the up-scattering of $\chi_1$ against electrons in the Sun to $\chi_2$\,\cite{An:2017ojc,An:2020bxd,Baryakhtar:2020rwy}. The $\chi_2$ that have outgoing velocities large enough to overcome the gravitational potential well of the Sun, can then reach the Earth (as depicted in fig.\,\ref{fig:ch4_upscatter}) and down-scatter at DD experiments via $\chi_2\,e\rightarrow\chi_1\,e$.
\\
\\
In the following, we follow the discussion in refs.\,\cite{Baryakhtar:2020rwy,An:2017ojc,Emken:2021lgc} to find the event rates at DD experiments. 
\subsection{Up-scattering from the Sun}
The maximum velocity of DM falling into the Sun is $\sqrt{(v^{\rm{esc}}_\odot)^2+(v_{\rm{DM}}^{\rm{\,halo}})^2}\simeq 4\times 10^{-3}$, where $v^{\rm{esc}}_\odot$ is the Solar escape velocity. We take the value of the Solar escape velocity to be the number averaged\footnote{$\langle v_{\rm esc}\rangle=\int_0^{r_{\rm core}}dr\,v_{\rm esc}(r)n(r)/\int_0^{r_{\rm core}}dr\,n(r)$ where $n(r)$ is the electron number density\,\cite{Emken:2021lgc} and $r_{\rm core}\simeq 0.2 R_\odot$.} escape velocity, over the Solar core\,\cite{Emken:2021lgc} (since most of the scatterings are expected to happen inside the core), giving $ v^{\rm{esc}}_\odot \simeq\, 1307\,\text{km/s}$\,\cite{Emken:2021lgc}. DM particle's most probable Galactic velocity\footnote{Using a halo distribution averaged velocity leads to an overall factor of $\simeq 0.89$ in the upscattered flux. } is given by $v_{DM}^{\textrm{halo}}\simeq220\,\text{km/s}$. This is much smaller than the most probable electron velocity in the Solar core
\begin{equation}
  v^e_{\odot}(\text{core})=\sqrt{\frac{2T_\odot}{m_e}}\simeq 0.066 \text{,  for } T_\odot=1.1\,\rm{keV},  
\end{equation}
where $T_\odot$ is the temperature at the Sun's core. Hence, the DM particles can be understood to be at rest, with Solar electrons scattering against them. The steady state DM number density in the Sun is given by
\begin{align}
n_{\chi_1,\odot}&=n_{\chi_1}^{\rm{\,halo}}\left(1+\left(\frac{v^{\text{esc}}_\odot}{v_{\text{DM}}^{\text{\,halo}}}\right)^2\right)\left(\frac{v_{\text{DM}}^{\text{\,halo}}}{v^{\rm{esc}}_\odot}\right) \nonumber\\
&\simeq n_{\chi_1}^{\rm{\,halo}}\left(\frac{v_{\rm{esc}}}{v_{\text{DM}}^{\text{\,halo}}}\right).
\end{align}

Here, the gravitational focusing effect,  that  enhances  the  area  at spatial infinity, is given by the factor $(1+(v^{\text{esc}}_\odot)^2/(v_{\text{DM}}^{\text{\,halo}})^2)$, and the factor $(v_{\text{DM}}^{\text{\,halo}}/v_{\rm{esc}})$ accounts for the decrease in the number density of DM owing to its larger velocity near the Sun (from conservation of flux).
The velocity distribution of electrons in the Sun is taken to be Maxwell Boltzmann\,(MB)
\begin{equation}
    f_{{\rm MB}}(v_e)=4\pi v_e^2\left(\frac{m_e}{2\pi T_\odot}\right)^{3/2}{\rm exp} \left(-\frac{m_e v_e^2}{2T_\odot}\right),
\end{equation}
where $m_e$ and $v_e$ are the electron mass and velocity, respectively.
The differential flux of particles $\chi_2$ generated with recoil energy $K_{\chi_2}$ is given as
\begin{eqnarray}
    \frac{d\Phi}{dK_{\chi_2}}&=&n_e \left\langle\frac{d\sigma_{\chi_1\rightarrow\chi_2} }{dK_{\chi_2}} v_e\right\rangle \frac{n_{\chi_1,\odot}V_\odot}{4\pi (1{\rm AU})^2}  \\
    &=& \frac{n_en_{\chi_1,\odot}V_\odot}{4\pi (1{\rm AU})^2}\int_{v_{min}(K_{\chi_2})}^{\infty}dv_e\, f_{MB}\left(v_e\right)\frac{d\sigma_{\chi_1\rightarrow\chi_2} }{dK_{\chi_2}}\,v_e \\
   &=&\begin{cases}
   \frac{n_en_{\chi_1,\odot}V_\odot}{4\pi (1{\rm AU})^2}
   \frac{\pi\alpha^2m_e^2}{\mu_{\chi e}^2}
   \frac{T_\odot}{m_e}
   \left(\frac{m_e}{2\pi T_\odot}\right)^{3/2}\bar{\sigma}_e\frac{1}{K_{\chi_2}}
   \text{exp}\left(-\frac{m_e}{4m_{\chi_1}T_\odot}\frac{(m_{\chi_1}K_{\chi_2}/\mu_{\chi e}+\delta)^2}{K_{\chi_2}}\right), \text{  EDM } \\
   \frac{n_en_{\chi_1,\odot}V_\odot}{4\pi (1{\rm AU})^2}\left(\frac{m_e}{2\pi T_\odot}\right)^{1/2}\bar{\sigma}_e^{F_{DM}(q)=1}\frac{m_{\chi_1}}{\mu_{\chi e}^2}
   \,\text{exp}\left(-\frac{m_e}{4m_{\chi_1}T_\odot}\frac{(m_{\chi_1}K_{\chi_2}/\mu_{\chi e}+\delta)^2}{K_{\chi_2}}\right) \\
   \qquad\times\Big(1+\frac{4m_{\chi_1}}{m_{\chi_1}-2m_e}\left(\frac{m_e^2}{4\mu_{\chi e}^2}+\frac{\delta m_e^2}{2\mu_{\chi e}m_{\chi_1}K_{\chi_2}}+\frac{m_eT_\odot}{m_{\chi_1}K_{\chi_2}}+\frac{m_e^2\delta^2}{4m_{\chi_1}^2K_{\chi_2}^2}\right)\Big),\text{  MDM,}
   \end{cases}\label{eq:dphidkchi2_prod}
\end{eqnarray}
where $n_e=2\times10^{25}\,\textrm{cm}^{-3}$ and $V_\odot=2.2\times 10^{31}\,\textrm{cm}^3$ are the Solar mean electron number density and volume, respectively\,\cite{Bahcall:2000nu,Baryakhtar:2020rwy}. 
The factor of $1/4\pi(1\textrm{AU})^2$ is the scaling in the flux on account of traveling from the Sun to the Earth, with the distance travelled being 1 AU (astronomical unit). The reduced mass is given by $\mu_{i,j}\equiv m_im_j/(m_i+m_j) $.

Here, we have plugged in the explicit expressions for the temperature averaged differential cross section for up-scattering, given by
\begin{equation}
    \left\langle  \frac{d\sigma_{\chi_1\rightarrow \chi_2}}{dK_{\chi_2}} v_e\right\rangle=\int_{v_{\rm min}}^\infty dv_e f_{\rm MB}(v_e) \frac{d\sigma_{\chi_1\rightarrow \chi_2}}{dK_{\chi_2}} v_e ,
\end{equation}
where  $v_{\rm min}(K_{\chi_2})$ is the minimum velocity an electron must have in order to up-scatter a $\chi_1$ into a $\chi_2$, with recoil energy $K_{\chi_2}$, given as
\begin{equation}
    v_{\rm min}=\frac{1}{\sqrt{2K_{\chi_2}m_{\chi_1}}}\left(\frac{m_{\chi_1}K_{\chi_2}}{\mu_{\chi_1,e}}+\delta\right),
\end{equation}
and 
\begin{equation}\label{eq:down_diff_cs}
    \frac{d\sigma_{\chi_1\rightarrow \chi_2}}{dK_{\chi_2}}\simeq \frac{\bar{\sigma}_e m_{\chi_1}}{2\mu_{\chi_1,e}^2v_e^2}|F_{DM}(q)|^2,
\end{equation}
 in the limit $\delta<<m_e$ and $m_{\chi_1}$.
Here, $\bar{\sigma}_e$ is the DM-electron reference cross section\,\cite{Essig:2011nj} with the 3-momentum
transfer fixed at $q=(\alpha m_e)$ (appropriate for atomic processes),
\begin{align}
    \bar{\sigma}_e&\equiv\frac{\mu_{\chi_1,e}^2}{16\pi m_{\chi_1}^2m_e^2}\overline{|\mathcal{M}_{\chi_1e}(q)|^2}\Big|_{q^2=(\alpha m_e)^2}\\
    &= \begin{cases}
            \frac{4d_\chi2}{\alpha}\frac{m_{\chi_1}^2}{(m_{\chi_1}+m_e)^2}, \text{ with  }F_{DM}(q)=\frac{\alpha m_e}{q}, &\text{ for EDM,} \\
                \left. \begin{array}{l}
                \alpha\mu_\chi^2\frac{m_{\chi_1}^2}{(m_{\chi_1}+m_e)^2}\left(1-\frac{2m_e}{m_{\chi_1}}\right),\text{ with } F_{DM}(q)=1, \\
                \frac{4}{\alpha} \mu_\chi^2 v^2  \frac{m_{\chi_1}^2}{(m_{\chi_1}+m_e)^2},\text{ with } F_{DM}(q)=\frac{\alpha m_e}{q}.
                \end{array}\right\} &\text{ for MDM. }
    \end{cases}\label{eq:sigmabare}
\end{align}
The $q$-dependence of the matrix elements are encoded in the DM form-factor $F_{DM}(q)$\,\cite{Essig:2011nj}.
$\overline{|\mathcal{M}_{\chi_1e}(q)|^2}$ is the squared-amplitude for DM–electron scattering, averaged (summed) over initial (final) spin states.
The MDM contribution consists of one part that is independent of relative velocity $v$ and with a form factor $F_{DM}(q)=1$ similar to that for contact interaction, and another part proportional to $v^2$ and with a form factor same as that of EDM.

\begin{figure}
    \centering
    \includegraphics[scale=0.5]{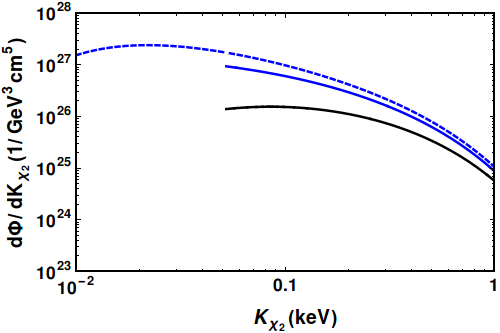}
    \caption{ The differential flux of particles $\chi_2$ generated with recoil energy $K_{\chi_2}$, $\frac{d\Phi}{dK_{\chi_2}}$, at production (blue-dashed), the flux that escapes the Sun's gravitational potential (blue-solid) and the attenuated flux at the Earth after taking into account $\chi_2$ decay in traveling from the Sun (black-solid).}
    \label{fig:dPhidKchi2}
\end{figure}
Note that the $\chi_2$ particles that reach the Earth are the ones that overcome the gravitational potential well of the Sun,
\begin{equation}\label{eq:grav_sun}
  v_{\chi_2}^2\simeq \frac{2K_{\chi_2}}{m_{\chi_2}}>(v^{\rm{esc}}_\odot)^2.  
\end{equation}
Therefore, the flux on the Earth is attenuated from the produced flux as
\begin{equation}
    \frac{d\Phi}{dK_{\chi_2}}(K_{\chi_2})\Bigg|_{Earth}=\frac{d\Phi}{dK_{\chi_2}}(K_{\chi_2}+m_{\chi_2}(v^{\rm esc}_\odot)^2/2)\;\Theta\left(K_{\chi_2}-\frac{1}{2}m_{\chi_2}(v^{\rm{esc}}_\odot)^2\right).\label{eq:flux_earth}
\end{equation}
The Heaviside theta function ensures that the condition in eq.\,\eqref{eq:grav_sun} is satisfied. We also shift the flux to account for the reduction in $\chi_2$ velocity in overcoming the Sun's gravitational potential. In the following, we denote the flux on the Earth by $d\Phi/dK_{\chi_2}$ and drop the explicit notation. There will be a further suppression in the flux of $\chi_2$ from its decay in the time taken to travel from the Sun to the Earth. In fig.\,\ref{fig:dPhidKchi2} we show the flux per unit energy from eq.\,\eqref{eq:dphidkchi2_prod} at production (blue-dashed), the flux that escapes the Sun's gravitational well from eq.\,\eqref{eq:flux_earth} (blue-solid) and the attenuated flux at the Earth after taking into account the $\chi_2$ decay in travelling from the Sun to the Earth (black-solid). Note that the flux that overcomes the gravitational potential has a lower cut-off, from the Heaviside theta function in eq.\,\eqref{eq:flux_earth}.

We note that the plasmon does not play any role in the up-scattering or production of DM in the Sun. The plasmon frequency in the non-relativistic limit\,\cite{Braaten:1993jw}
\begin{equation}
    \omega_p^2(T)=\frac{4\pi\alpha\, n_e}{m_e}\left(1-\frac{5}{2}\frac{T}{m_e}\right)
\end{equation}
gives a plasmon mass order 1 smaller than the Sun's average temperature for the Solar electron number density. Therefore, the plasmon effect is subdominant in scattering. In addition, there can be no plasmon-sourced production as we consider DM masses much larger than the Solar temperature. 

\subsection{Down-scattering on the Earth}
\subsubsection{Electron scattering:}
The $\chi_2$ flux received on the Earth is depleted if the decay lifetime of $\chi_2$ is comparable to the time taken by it to travel to the Earth. We take this into account, and calculate the down-scattering event rate in electron scattering experiments. 
With the DM flux per unit energy $d\Phi/dK_{\chi_2}$ given in eq.\,\eqref{eq:flux_earth}, we can write the electron recoil energy spectrum per detector mass per unit time as\,\cite{Essig:2015cda,Bloch:2020uzh,Baryakhtar:2020rwy}
\begin{eqnarray}
    \frac{dR_{\rm ion}}{d\Delta E_e}&=&n_T\,  \epsilon(\Delta E_e)\sum_{n,l}\frac{1}{\Delta E_e-E_{nl}}
    \frac{\bar{\sigma}_e}{64\mu_{\chi_2,e}^2}
    \int dK_{\chi_2}\Theta(\Delta E_e^{\rm max}(K_{\chi_2})-\Delta E_e)\frac{d\Phi}{dK_{\chi_2}}\frac{m_{\chi_2}}{K_{\chi_2}}
    e^{-t(K_{\chi_2})\times \Gamma_{\chi_2}}\nonumber\\
    & &\qquad\qquad\times \int_{q^-(K_{\chi_2},\Delta E_e,\delta,m_{\chi_2})}^{q^+(K_{\chi_2},\Delta E_e,\delta,m_{\chi_2})} dq\,q|F_{DM}(q)|^2|f_{nl\rightarrow \Delta E_e-E_{nl}}(q)|^2  ,
    \label{eq:e_rec_spec}
\end{eqnarray}
where $\Delta E_e$ is the energy transferred to the electron\footnote{The energy transferred to the electron is a sum of the energy of the outgoing electron at asymptotically large distances from the nucleus, $E_R$, and the ionization energy of the shell it originated from, $E_{nl}$, i.e., $\Delta E_e\equiv E_R+E_{nl}$. The energies are assumed to be emitted almost simultaneously and the collection of the energies of the electrons and photons emitted at the de-excitation and the ionization together is assumed to be equal to $\Delta E_e$, that is we assume that the ionization energy is released completely\,\cite{Ibe:2017yqa}.}, $E_{nl}$ is the ionization energy of the $n,\,l$ orbital of given atom, $\epsilon(\Delta E_e)$ is the signal efficiency given in fig.\,2 of \,\cite{XENON:2020rca}. The maximum energy transferred to the electron for a given incoming DM kinetic energy $K_{\chi_2}$ is given by $\Delta E_e^{\rm max}(K_{\chi_2})=K_{\chi_2}+\delta$. For DM-electron scattering in xenon, we get the number of targets per tonne as $n_T=2\times 4.2\times 10^{27}/\text{tonne}$\footnote{We take $Z_{\rm eff}=2$ since the electrons in different orbitals are accounted for by summing over ionization form factors for all the accessible orbitals}. The exponential factor accounts for the depletion in $\chi_2$ flux in travelling from the Sun to the Earth and $t(K_{\chi_2})= 1{\rm AU}\times\sqrt{m_{\chi_2}/2K_{\chi_2}}$ is the time taken by $\chi_2$ to travel to the Earth. The decay width of $\chi_2$ is,
\begin{equation}\label{eq:chi2_decay}
    \Gamma_{\chi_2}\simeq (d_\chi,\mu_\chi)^2\delta^3/\pi.
\end{equation}
The form factor for ionization of an electron in $n,l$ orbital with a total of $\Delta E_e$ energy transferred to the electron, is given by  $|f_{nl\rightarrow \Delta E_e-E_{nl}}(q)\big|^2$. We use {\tt QEDark}\,\cite{Essig:2015cda} to extract these ionization form factors.

The limits of integration $q^\pm$ are obtained from energy conservation in the down-scattering process and given as 
\begin{equation}
    \frac{q^2}{2m_{\chi_2}}-v\,q\,\textrm{cos}\,\theta=\delta-\Delta E_e,
\end{equation}
where $\theta$ is the angle between the momentum of $\chi_2$ and the transferred momentum $q$ and
\begin{equation}
    |\textrm{cos}\,\theta|\leq 1\implies q^\pm(v,\Delta E_e,\delta,m_{\chi_2})=\Big|m_{\chi_2}v\pm\sqrt{m_{\chi_2}^2v^2-2m_{\chi_2}(\Delta E_e-\delta)}\Big|.
\end{equation}
We note that the factor from integration over transferred momentum $q$ in eq.\,\eqref{eq:e_rec_spec}
\begin{equation*}
    f_{ion}^{int}(\Delta E_e,q)=\int_{q^-(v,\Delta E_e,\delta,m_{\chi_2})}^{q^+(v,\Delta E_e,\delta,m_{\chi_2})} dq\,q\frac{\alpha^2 m_e}{4(\Delta E_e-E_{nl})}|F_{DM}(q)|^2\big|f_{nl\rightarrow \Delta E_e-E_{nl}}(q)\big|^2,
\end{equation*}
leads to an enhancement in the $F_{DM}(q)=1$ part of the MDM scattering rate but a small suppression for the remaining part of MDM as well as the EDM scattering rate.


We find the limits from latest results from XENONnT experiment\,\cite{XENONCollaboration:2022kmb} with a total exposure of $1.16\,\textrm{tonne}\times\textrm{year}$ and background rate of $15.8\pm 1.3$ \,events$/(\textrm{tonne}\times\textrm{year}\times\textrm{keV})$.
We also find the projected limits from future runs of XENONnT and DARWIN experiments, with projected exposures of $20\,{\rm tonne}\times {\rm year}$ and $200\,{\rm tonne}\times {\rm year}$, respectively. We use the same background rate for future projections, as that of the latest XENONnT run, to get the most conservative limits.
We also use the same efficiency as of XENON1T, which gives a conservative bound, since the efficiency can only be expected to improve. 
We discuss the results in section\,\ref{sec_res}.
\subsubsection{Scattering from Migdal effect:}
In addition to the direct ionization of an electron, a DM particle scattering off of a nucleus can also lead to subleading electronic energy deposition into detectors via \textit{nuclear} scattering. In a typical nuclear recoil the electron cloud is assumed to follow the recoiling nucleus instantaneously. But if the effect of the sudden acceleration of the nucleus, with the electron cloud still in its original position, is taken into account, it is known to deposit electronic energy via ionization/excitation of the recoiling atom (the Migdal effect) or the emission of a Bremsstrahlung photon\,\cite{migdal1941ionization,Bernabei:2007jz,Ibe:2017yqa,XENON:2019zpr}.

In the Migdal approximation, the whole electron cloud is assumed to recoil with the same  velocity with respect to the nucleus, with no change in its shape. The rate for a nuclear scattering with recoil energy $E_{\rm NR}$ 
accompanied with a Migdal electron recoil with energy $E_e$ from $n,l$ orbital, 
leading to a deposition of total energy $\Delta E_e\simeq E_e+E_{nl}$, is\,\cite{Ibe:2017yqa,Dolan:2017xbu,Bell:2021zkr}
\begin{eqnarray}
\frac{d^3R}{dE_{\rm NR}dE_edK_{\chi_2}}=\sum_{n,l}\frac{d^2R}{dE_{\rm NR}dK_{\chi_2}}\big|Z(E_{\rm det}-\mathcal{L}E_{\rm NR}-E_{nl})\big|^2,
\end{eqnarray}
where, $E_{\rm det}=E_e+E_{nl}+\mathcal{L}E_{\rm NR}$. The Lindhard quenching factor is denoted by $\mathcal{L}$  and is the fraction of the nuclear recoil energy observed in the electron channel. Its value is well approximated to $\mathcal{L}\simeq 0.15$ \cite{Bell:2021zkr,Essig:2015cda}. 
The ionization form factor is given by $\big|Z(E_{\rm det}-\mathcal{L}E_{\rm NR}-E_{nl})\big|^2$ and its values for different orbitals are given in fig.\,4 of ref.\,\cite{Ibe:2017yqa}.
The nuclear differential scattering rate is:
\begin{eqnarray}
    \frac{d^2R}{dE_{\rm NR}dK_{\chi_2}}=n_T\frac{d\Phi}{dK_{\chi_2}}\frac{d\sigma_N}{dE_{\rm NR}}e^{-t(K_{\chi_2})\times \Gamma_{\chi_2}}.
\end{eqnarray}
The nuclear differential scattering cross sections, approximated to the elastic case, are
\begin{equation}
    \frac{d\sigma_N}{dE_{\rm NR}}\simeq 
    \begin{cases}
    Z^2\alpha \frac{d_\chi^2}{16}\frac{1}{v^2E_{\rm NR}}, & \textrm{for EDM,} \\
    Z^2\alpha \frac{\mu_\chi^2}{16}\frac{1}{E_{\rm NR}}\left(1+\frac{E_{\rm NR}}{2\mu_{\chi_2,N}v^2}\right), & \textrm{for MDM Spin Independent,}\\
    \alpha \frac{\mu_\chi^2}{16}\left(\frac{\mu_{Z,N}}{e/2m_N}\right)^2\frac{m_N}{m_n^2v^2}, & \textrm{for MDM Spin Dependent},
    \end{cases}\label{eq:dsigder}
\end{equation}
where $m_N$ and $Z$ are the mass and atomic number of the nucleus, respectively, and $E_{\rm NR}$ is the nuclear recoil energy. 

The most sensitive low-energy analysis comes from the S2-only data set from the XENON1T experiment\,\cite{XENON:2019gfn}. The S2-only differential rate is
\begin{equation}
    \frac{dR}{dE_{\rm det}}=\int dE_{\rm NR}\int_{K_{\chi_2}^{min}}^\infty dK_{\chi_2} 
    \frac{d^3R}{dE_{\rm NR}dE_edK_{\chi_2}},
    \label{migdal_rate}
\end{equation}
where the minimum kinetic energy of the incoming DM $\chi_2$ to downscatter to $\chi_1$, with a nuclear recoil energy of $E_{\rm NR}$ along with an electronic deposition of energy $E_{\rm det}$ via Migdal effect, is given by
\begin{eqnarray}
    K_{\chi_2}^{min}&=&\frac{m_{\chi_2}}{2}(v_{min}^{\rm Mig,inel})^2, \\
    {\rm where,\,} v_{min}^{\rm Mig,inel}&=&\sqrt{\frac{m_NE_{\rm NR}}{2\mu_{\chi_2,N}^2}}+\frac{\Delta E_e-\delta}{\sqrt{2m_N E_{\rm NR}}},\\
    &=& \sqrt{\frac{m_NE_{\rm NR}}{2\mu_{\chi_2,N}^2}}+\frac{E_{\rm det}-\mathcal{L}E_{\rm NR}-\delta}{\sqrt{2m_N E_{\rm NR}}}.
\end{eqnarray}
We find the total number of event at XENON1T  by integrating  eq.\,\eqref{migdal_rate} over the range $E_{\rm det}=0.19-3.8$\,keV$_{ee}$, with an exposure of 22 tonne-day\,\cite{XENON:2019gfn}. A total of 61 events were observed at XENON1T over this exposure, while the expected number of background events  was 23.4. This gives an upper limit of 49 events expected from DM at 90$\%$ confidence, and can be used as an upper limit to derive constraints on DM interactions with SM. For our model of dipolar DM though we find less than 1 events over the full parameter space of interest, so the scattering rate from Migdal effect is not large enough to derive any bounds on Solar upscattered dipolar DM.\\
We note that this is expected from the discussion in \cite{Essig:2019xkx} for long range interactions, like EDM $F_{DM}(q)=1/q^2$, with a bias towards low $q$ values where the Migdal effect rates are always smaller than electron ionization rates by a factor of $\sim Z^2m_e^2/m_N^2$. 

\section{Complementary Bounds on sub-GeV Dark Matter}\label{sec:CB}
In this section, we discuss the constraints on sub-GeV DM with $\mathcal{O}$(keV) splittings from existing laboratory experiments and astrophysical sources.
\subsection{Fixed target experiments}
\begin{figure}[b!]
    \centering
    \begin{subfigure}[b]{0.33\textwidth}
    \includegraphics[width =\textwidth]{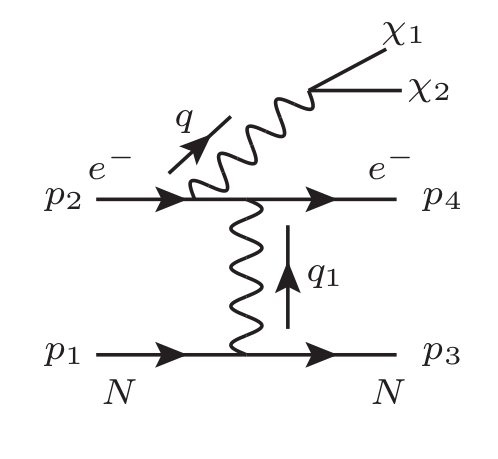}
    \caption{}
    \label{fig:NA64_ISR}
    \end{subfigure}
    \begin{subfigure}[b]{0.385\textwidth}
    \includegraphics[width=\textwidth]{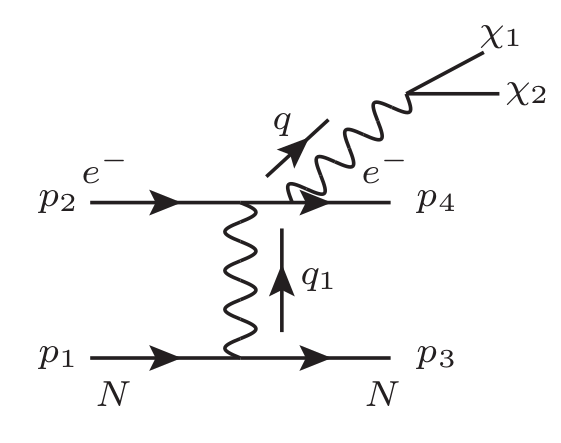}
    \caption{}
        \label{fig:NA64_FSR}
    \end{subfigure}
\caption{Leading processes for production of dark sector particles via off-shell photon in a fixed target experiment.}
\label{fig:NA64}
\end{figure}
At lepton-fixed target experiments, an electron beam of fixed energy is dumped against an active target, comprised of some heavy nucleus, that is either a part of the detector itself or a separate target.
Dark sector particles can be pair produced from the electrons scattering off of nucleons, giving rise to missing-energy final states\,\cite{Gninenko:2013rka,Andreas:2013lya}. These experiments employ stringent selection criteria making it possible to conduct an essentially background free search for such missing-energy signals\,\cite{Bjorken:2009mm,Tsai:1973py, Graham:2021ggy,Fabbrichesi:2020wbt,Agrawal:2021dbo}.

In particular, we use results from the NA64 experiment at CERN SPS\,\cite{NA64:2017vtt} to derive constraints on our model via the production of DM particles from an off-shell photon,
\[e^-N\rightarrow e^-N\gamma_{\rm vir}\rightarrow e^-N\chi_1\chi_2,\]
where $\gamma_{\rm vir}$ is the off-shell photon that subsequently produces the dark sector particles. The leading processes are shown in fig.\,\ref{fig:NA64}. Note that we do not consider the processes where the virtual photon originates from the nucleus since these processes are suppressed by a factor of $(Zm_e/m_N)^2$ for coherent photon emission. 
We follow the discussion in ref.\,\cite{Chu:2018qrm} in the following. 

The NA64
experiment employs the optimized 100\,GeV electron beam from the H4  beamline at the North Area\,(NA) of the CERN SPS. 
The beam is incident upon an electromagnetic calorimeter\,(ECAL) made up of a $6\times 6$ matrix with Pb and Sc plates, each module being $\simeq 40$ 
radiation lengths ($X_0$) long. 
The radiation length\footnote{For incident electrons with large energies, this is essentially a measure of the strength of the Bremsstrahlung process with a larger radiation length implying smaller cross sections for the process.} of Sc is about 1 order larger than that of Pb so $e^-$ scattering with Sc is subdominant. \\
Assuming that the fermion pair is produced within the first radiation length gives the target length to be $L_{\textrm{target}}=X_0=0.56\,$cm for Pb.
The search region in the ECAL is limited by the energy  threshold for detection of electron on one side and the requirement for missing energy to be larger than half the beam energy. This gives the selection criteria for the energy of the outgoing electron  $E_4$ as\,\cite{Banerjee:2019pds,Krasnikov:2020trl}
\begin{equation}
 0.3\,\textrm{GeV}<E_4<50\,\textrm{GeV}  .\label{eq:NA64_E4} 
\end{equation}
The polar angular coverage for the outgoing electron is
\begin{equation}
  0.0<\theta_4<0.23\,\,{\rm radians}.  \label{eq:NA64_theta4}
\end{equation}
The number of signal events with these geometric and angular cuts are\,\cite{Chu:2018qrm}
\begin{equation}\label{eq:NA64_events}
    N_{\textrm{sig}}=N_{\textrm{EOT}}\frac{\rho_{\textrm{target}}}{m_N}L_{\textrm{target}}\int_{E_4^{min}}^{E_4^{max}}\epsilon_{eff}(E_4)\int_{\textrm{cos}\,\theta_4^{min}}^{\textrm{cos}\,\theta_4^{max}}d\textrm{cos}\,\theta_4\frac{d\sigma_{\textrm{prod}}}{dE_4d\textrm{cos}\,\theta_4},
\end{equation}
where, $N_{\textrm{EOT}}=2.84\times 10^{11}$ are the number of electrons incident upon the target\,\cite{Banerjee:2019pds}. The target material density and nuclear mass are given by $\rho_{\rm target}$ and  $m_N$, respectively. The detector efficiency of NA64 is known to depend only marginally on the energy and is taken to be constant, $\epsilon_{eff}(E_4)\simeq 0.5$ (averaging over the total signal efficiencies of the various runs\,\cite{Banerjee:2019pds}). The integration limits of $E_4$ and $\theta_4$ are from eqs.\,\eqref{eq:NA64_E4} and \eqref{eq:NA64_theta4}, respectively. The double differential cross section for production of the processes shown in figs.\,\ref{fig:NA64_ISR} and \ref{fig:NA64_FSR} is given by $d\sigma_{prod}/dE_4d\cos\theta_4$ with the full expression given in eq.\,\eqref{app4:NA64_diffsigma}.

Since the beam energy is much larger than the $\mathcal{O}$(keV) splitting we consider, and the signal being observed is of missing energy in final state, the constraints on our model do not differ from those of the elastic DM cases discussed in ref.\,\cite{Chu:2018qrm}. 
We follow the discussions in\,\cite{Chu:2018qrm} and re-derive the constraints from their run as mentioned in ref.\,\cite{Banerjee:2019pds}. 
The details of the calculation are given in Appendix\,\ref{app4:NA64}.
We derive constraints by demanding that $N_{sig}<N^{90\%}=2.3$ where the latter corresponds to the $90\%$ C.L. for the number of signals events given zero observed events. The resulting constraints are shown in figs.\,\ref{fig:ch4_EDM_res} and \ref{fig:ch4_mdm_res}.

This type of search for missing-energy in the final state gives stronger bounds for a feebly interacting dark particle, as compared to experiments where the dark particle is detected via its scattering off electrons/nuclei in the main detector (for example in the mQ experiment at SLAC\,\cite{Prinz:1998ua}), since processes of the latter kind are suppressed by further powers of the small-valued DM-SM effective coupling. We, therefore, do not study these latter processes. 

Constraints are also derived from proton fixed target experiments, and as shown in ref.\,\cite{Chu:2020ysb} the strongest such constraint\footnote{Constraints coming from other proton-beam experiments such as COHERENT, JSNS$^2$, NO$\nu$A and WA66 are expected to be weaker than that from CHARM-II and LEP\,\cite{Chu:2020ysb}.} comes from  CHARM-II experiment, which used a 450 GeV proton beam on a Be target. 
The constraints are derived from single electron recoil events at recoil energies $E_R \in [3,24]$\,GeV, so that the small splitting of $\mathcal{O}$(keV) has negligible effect and the constraints for such inelastic DM are the same
as those for the elastic case. These are shown in figs.\,\ref{fig:ch4_EDM_res} and \ref{fig:ch4_mdm_res}.

\subsection{Production at lepton colliders}
\begin{figure}[b!]
\centering
\begin{subfigure}[b]{0.4\textwidth}
    \includegraphics[width=\textwidth]{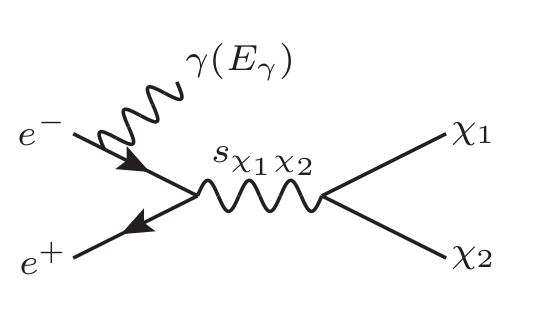}
\end{subfigure}
\begin{subfigure}[b]{0.4\textwidth}
    \includegraphics[width=\textwidth]{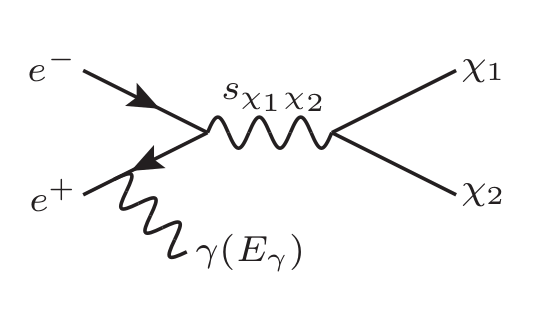}
\end{subfigure}
\caption{Processes for production of DM along with initial state radiation, at $e^+e^-$ colliders, leading to missing energy plus mono-photon signatures.}
\label{fig:babar2}
\end{figure}

DM can be produced from $e^+e^-$ collisions at lepton colliders and appear as missing energy\,($\slashed{E}$), since they do not scatter within the collider. 
Along with initial state radiation\,(ISR) or final state radiation\,(FSR), this leads to a particularly clean signature of mono-photon plus missing energy\,($\gamma+\slashed{E}$).
The FSR processes are suppressed by the small DM-photon couplings, so we only consider the ISR process, as shown in fig.\,\ref{fig:babar2}, for deriving constraints.

We follow the discussion in ref.\,\cite{Chu:2019rok} for constraints on DM dipole moments from pair production, adapting them for the inelastic case. For DM mass splittings much smaller than the CM energy of the $\chi_1,\chi_2$ system, $\delta\sim\mathcal{O}(\text{keV})<<\sqrt{s_{\chi_1\chi_2}}$, these constraints for inelastic DM are equal to the ones for elastic DM.
\subsubsection{BABAR}\label{babar}
We use the data from the search for mono-photon events in decays of $\Upsilon(3S)$
\begin{equation*}
  \Upsilon(3S)\rightarrow \gamma (A^0\rightarrow inv.)  
\end{equation*}
at BABAR detector at the PEP-II asymmetric-energy $e^+e^-$ collider at the Stanford Linear Accelerator Center (SLAC), with 28\,fb$^{-1}$ of data collected at a CM energy\,\cite{BaBar:2008aby}
\begin{equation*}
  \sqrt{s}\approx m_{\Upsilon(3S)}\approx\,10.355\,\textrm{GeV.}  
\end{equation*}
The single-photon events were chosen based one two trigger criteria,
\begin{align}
    \textrm{High-E region:\,\,}& 3.2<E_{\gamma}<5.5\,\textrm{GeV,\,}-0.31< \textrm{cos}\,\theta_{\gamma}<0.6\,\,{\rm radian},&\textrm{full data},\nonumber\\
    \textrm{Low-E region:\,\,}& 2.2<E_{\gamma}<3.7\,\textrm{GeV,\,}-0.46< \textrm{cos}\,\theta_{\gamma}<0.46\,\,{\rm radian},&\textrm{19\,fb$^{-1}$ of data},\nonumber 
\end{align}
where $\theta_\gamma$ is the CM polar angle, and $E_{\gamma}$ is the photon energy in the $\Upsilon$ rest frame.
For each region, the number of signal events is given by
\begin{equation}
    N_{sig}^{(i)}=\epsilon_{\text{eff}}\,\mathcal{L}\int_{\text{bin }i} \frac{ds_{\chi_1\chi_2}}{s}\int_{\textrm{cos}\,\theta_\gamma^{\text{min}}}^{\textrm{cos}\,\theta_\gamma^{\text{max}}}d \,\textrm{cos}\,\theta_\gamma \frac{d\sigma_{e^+e^-\rightarrow \chi_1\chi_2\gamma}}{dx_\gamma d\textrm{cos}\,\theta_\gamma}.
\end{equation}
Here, $\epsilon_{\text{eff}}$ is the total efficiency, $\mathcal{L}$ is the integrated luminosity, $\sqrt{s}$ is the CM energy of the $e^+e^-$ system, $\sqrt{s_{\chi_1\chi_2}}\equiv \sqrt{(1-x_\gamma)s}$ is the CM energy of the $\chi_1\chi_2$ system with $x_\gamma=E_\gamma/E_{beam}=2E_\gamma\big/\sqrt{s}$. The photon makes an angle of $\theta_\gamma$ with respect to the beam in the CM frame. 
Following ref.\,\cite{Essig:2013vha} we apply a non-geometric cut $\epsilon_{\text{eff}}$ of $30\%$ and $55\%$ in the high-E and low-E regions, respectively. 

The ISR production cross section is approximated by dressing the cross section for DM pair production (without ISR), with an angle-dependent radiator function\,\cite{Montagna:1995wp,Chu:2019rok} as:
\begin{equation}
    \frac{d\sigma_{e^+e^-\rightarrow \chi_1\chi_2\gamma}}{dx_\gamma d\textrm{cos}\,\theta_\gamma}=\sigma_{e^+e^-\rightarrow \chi_1\chi_2}(s,s_{\chi_1\chi_2})\mathcal{H}^{(\alpha)}(x_\gamma,\theta_\gamma,s),
\end{equation}
where the cross-section for $e^+e^-\rightarrow \chi_1\chi_2$ at the energy scale reduced by photon emission is
\begin{align}
   \sigma_{e^+e^-\rightarrow \chi_1\chi_2}
    &=\frac{1}{32\pi \sqrt{s \,s_{\chi_1\chi_2}}}\frac{|\vec{p}_f|}{|\vec{p}_i|}\int d(\textrm{cos}\,\theta)\, \overline{|\mathcal{M}|^2}    \\
    &=\frac{\alpha(s+2m_e^2)}{6s_{\chi_1\chi_2}^3}
    \sqrt{\frac{m_{\chi_1}^4-2m_{\chi_1}^2(m_{\chi_2}^2+s_{\chi_1 \chi_2})+(m_{\chi_2}^2-s_{\chi_1\chi_2})^2}{s(s-4m_e^2)}}\nonumber\\
     &\quad\times
    \begin{cases}
    d_\chi^2 \Big(s_{\chi_1\chi_2}^2+s_{\chi_1\chi_2}(m_{\chi_1}^2-6m_{\chi_1}m_{\chi_2}+m_{\chi_2}^2)-2\,(m_{\chi_2}^2-m_{\chi_1}^2)^2\Big) , & \text{for EDM}, \\
    \mu_\chi^2 \Big(s_{\chi_1\chi_2}^2+s_{\chi_1\chi_2}(m_{\chi_1}^2+6m_{\chi_1}m_{\chi_2}+m_{\chi_2}^2)-2\,(m_{\chi_2}^2-m_{\chi_1}^2)^2\Big) 
     , & \text{for MDM}, \\
\end{cases}       
\end{align}
and, the radiator function taking into account all soft and collinear corrections upto $\mathcal{O}(m_e^2/s)$ is \,\cite{Montagna:1995wp}
\begin{equation}
    H^{\alpha}(x_\gamma,\theta_\gamma,s)=\frac{\alpha}{\pi}\frac{1}{x_\gamma}\left[\frac{1+(1-x_\gamma)^2}{1+4m_e^2/s-\textrm{cos}^2\theta_\gamma}-\frac{x_\gamma^2}{2}\right].
\end{equation}
To constrain the DM couplings, we require that the expected number of events be smaller than the observed number of events in each bin $i$ at 90$\%\,$C.L., such that $N_{sig}^{(i)}<N_{obs}^{(i)}+1.28\,\sigma_{obs}^{(i)}$. We show these bounds in figs.\,\ref{fig:ch4_EDM_res} and \ref{fig:ch4_mdm_res}.
\subsubsection{LEP}
We consider high energy colliders like the Large Electron Positron (LEP) collider. The model considered here can give rise to events with one photon and missing energy by the same process as at BABAR, see section\,\ref{babar}. The only difference lies in the high CM energies $189\,\textrm{GeV}\leq\sqrt{s}\leq209\,$GeV and high luminosities (a total of $619$\,pb$^{-1}$ of data for the single- and multi-photon $+$ missing energy final states) that the LEP operated at\,\cite{L3:2003yon}. 
The cross sections for these processes having been found to be in agreement with the SM expectation from $e^+e^-\rightarrow \nu \bar{\nu}\gamma(\gamma)$ give constraints on any BSM physics model that can also lead to the same final state. But the high CM energies lead to constraints that are independent of the DM mass, for light DM ($m_\chi<$\,GeV here). This was studied in ref.\,\cite{Fortin:2011hv} and bounds obtained from the mono photon channels give
\begin{equation}
    |\mu_\chi|\textrm{\,or\,} |d_\chi|<1.3\times10^{-5}\mu_B,
\end{equation}
where $\mu_B\equiv e/2m_e$ is the Bohr magneton. We note that at these large energies of production, the small splitting of $\mathcal{O}$(keV) has negligible effect and the constraints for such inelastic DM are the same as those for the elastic case. The upper bound is shown in figs.\,\ref{fig:ch4_EDM_res} and \ref{fig:ch4_mdm_res}. 

Note that the LHC probes heavier masses and smaller couplings $\sim\mathcal{O}(0.01)$\,GeV$^{-1}$\,\cite{Barger:2012pf}, and is not of relevance in this work.
\subsection{Supernovae cooling}
Light DM, $m_{DM}\leq \mathcal{O}(100\textrm{\,MeV}$), can be produced in supernovae with core temperatures of $\mathcal{O}$(30\,MeV)\,\cite{Raffelt:1996wa,Dreiner:2003wh,Fischer:2016cyd,Magill:2018jla,Chang:2016ntp,Chang:2018rso}. If these dark particles escape, they can cause extra cooling and lead to changes in the shape and duration of the neutrino pulse. 
Light dark particles can thus be constrained by comparing neutrino pulse predictions to those observed from the SN1987A at terrestrial neutrino observatories\,\cite{Kamiokande-II:1987idp,Bionta:1987qt,Alekseev:1988gp}, assuming that the SN1987A was a neutrino-driven supernova (SN) explosion\,\cite{Chu:2019rok}. 
The constraints derived from SN depend on their cooling rate, with the predominant process being
\begin{equation}\label{eq:SN_process}
    e^+(p_1),e^-(p_2)\rightarrow\chi_1(p_3),\chi_2(p_4)
\end{equation}
since positrons are thermally supported in SN. 
Constraints on models are derived in two limiting cases leading to two-sided bounds on DM couplings for each mass as follows:
\begin{enumerate}
    \item Weak coupling: This is applicable in the limit of small interaction strengths of DM with SM such that for any smaller strengths there would be too little production of DM to cause any significant change to the SN cooling rate. In this limit, any DM that is produced escapes the SN with almost $100\%$ probability, such that it is possible to derive lower bounds on the DM effective coupling, by constraining only the production rates. This is given by the 
``Raffelt criterion"\,\cite{Raffelt:1996wa} which says that any ``exotic" cooling will not change the neutrino signal significantly, as long as the emissivity obeys the condition
\begin{equation}
    \dot{\mathcal{E}}<10^{19}\,\textrm{erg/g/s}.
\end{equation}
This condition is easily converted into a condition on energy emitted per unit time per unit volume by noting that the density for $t\simeq 1\textrm{s}$ is nearly constant (see fig.\,5 of ref.\,\cite{Burrows:1986me}), giving a conversion between $d^3r$ and $dM$. We take $\rho\approx 3\times 10^{14}\, \textrm{g/cm}^3$.
The emissivity (energy emitted per unit volume per unit time) is defined as \cite{Dreiner:2003wh}
\begin{align}
    \frac{d\mathcal{E}}{dt}&=\int d\Pi_{i=1,4} \frac{d^3p_i}{(2\pi)^3 (2E_i)}(2\pi)^4\delta^4(p_1+p_2-p_3-p_4)f_1f_2(1-f_3)(1-f_4)|\mathcal{M}|^2(E_3+E_4),\label{inel:emmissivity}
\end{align}
where $f_i$ are the Fermi-Dirac distribution functions
\begin{equation}
    f_i\equiv \frac{1}{\textrm{exp}\left(\frac{E_i-\mu_i}{T}\right)+1}.
\end{equation}
We ignore the final state Pauli blocking for small DM number densities and simplify the expression using
\begin{align}
    &\int d\Pi_{i=1,2} \frac{d^3p_i}{(2\pi)^3 (2E_i)} (2\pi)^4\delta^4(p_1+p_2-p_3-p_4)f_1f_2(1-f_3)(1-f_4)|\mathcal{M}|^2(E_3+E_4)\nonumber \\
    & \quad\simeq (E_1+E_2)f_1f_2\times4v_{m\textrm{\o}l}E_1E_2\sigma'\\
    &\quad =(E_1+E_2)f_1f_2\times 4F\sigma',
\end{align}    
where we use energy conservation $E_1+E_2=E_3+E_4$ and $F\equiv\left( \sqrt{s(s-4m_e^2)}/2\right)$ \cite{Gondolo:1990dk}.
The cross-section of production for the process in eq.\,\eqref{eq:SN_process} with squared-amplitude \textit{summed} over initial and final spins\footnote{$\sigma'=g_1g_2\sigma$ where $g_1,g_2$ are the spin degrees of freedom of incoming SM fermions, and $\sigma$ is the usual cross-section defined with the squared-amplitude averaged(summed) over initial(final) spins.},
is given in the limit of vanishing electron mass $m_e^2<s\approx T_{avg}^2\simeq(30\,\textrm{MeV})^2$ as
\begin{align}
\sigma'&=\frac{2 \alpha}{3s^2}
   \sqrt{\frac{m_{\chi_1}^4-2 m_{\chi_1}^2
   \left(m_{\chi_2}^2+s\right)+\left(m_{\chi_2}^2-s\right)^2}{s^2}}\nonumber\\
   &\qquad\times
   \begin{cases}
   d_\chi^2\left(s (m_{\chi_1}^2-6 m_{\chi_1} m_{\chi_2}+m_{\chi_2}^2)-2 (m_{\chi_1}^2-m_{\chi_2}^2)^2+s^2\right), & \text{for EDM},\\
   \mu_\chi^2\left(s (m_{\chi_1}^2+6 m_{\chi_1} m_{\chi_2}+m_{\chi_2}^2)-2 (m_{\chi_1}^2-m_{\chi_2}^2)^2+s^2\right), & \text{for MDM.} \\
   \end{cases}\label{inel:SN_prod}
\end{align}
Also simplifying the remaining part of eq.\,\eqref{inel:emmissivity} as\,\cite{Gondolo:1990dk,Dutra:2019gqz}
\begin{equation}
    \int \frac{d^3p_1}{(2\pi)^3 (2E_1)}\frac{d^3p_2}{(2\pi)^3 (2E_2)}=\frac{1}{8(2\pi)^4}\int ds \int_{\sqrt{s}}^\infty dE_+\,E_+ \int_{-\sqrt{E_+^2-s}}^{\sqrt{E_+^2-s}}dE_-,
\end{equation}
with $E_\pm\equiv E_1\pm E_2$, we rewrite eq.\,\eqref{inel:emmissivity} in the limit of vanishing electron mass as
\begin{align}
    \frac{d\mathcal{E}}{dt}=\frac{1}{4(2\pi)^4}\int_{(m_{\chi_1}+m_{\chi_2})^2}^\infty ds\,s\,\sigma'(s)\int_{\sqrt{s}}^{\infty}dE_+ E_+\int_{E_-^{min}}^{E_-^{max}}dE_- f_1f_2
\end{align}
where 
\[f_{1,2}=\frac{1}{\textrm{exp}\left(\frac{E_+\pm E_-\mp 2\mu_e(r_0)}{2T(r_0)}\right)+1},\]
computed at radius $r_0=10$\,km, where the  emmissivity can be seen to be maximum. We use the temperature and chemical potential radial profiles as given in ref.\,\cite{Magill:2018jla}.  The limits of $E_-$ integration are $E_-^{max,min}\equiv \pm\sqrt{1-4m_e^2/s}\sqrt{E_+^2-s}$.
\item Large coupling:
In the opposite limit of large DM couplings, the cooling process is dictated by the probability of escape or mean free path\,(MFP) of DM. The relatively larger density of SN results in a trapping of DM particles produced inside the SN, giving an upper bound on the DM effective coupling with SM. We adapt this bound from ref.\,\cite{Chu:2018qrm}, shown in figs.\,\ref{fig:ch4_EDM_res} and \ref{fig:ch4_mdm_res}.
\end{enumerate}

\section{Results}\label{sec_res}
We summarise the results for sub-GeV pseudo-Dirac DM with mass states $\chi_1$ and $\chi_2$ having mass difference $\mathcal{O}$(keV) for the two interactions:
\begin{enumerate}
    \item Transition EDM interaction:
    \begin{itemize}
        \item As discussed in section\,\ref{sec_relic}, the FI production is UV sensitive. We show the relic density contours for two reheating temperatures, 5\,MeV and 10\,MeV, in fig.\,\ref{fig:ch4_EDM_res}. For higher values of reheating temperatures, the contours would shift to lower $d_\chi$ values and the upward bend would occur at higher $m_{\chi_1}$, going outside the range shown here.  The $T_{RH}=5\,{\rm MeV}$ contour can be seen to cut off at $m_{\chi_1}\simeq 100\,{\rm MeV}$ as the observed relic density cannot be reproduced via FI production for larger masses. We show the relic contours for two values of mass splittings, $\delta=1\,{\rm keV}$ (dashed) and $\delta=10\,{\rm keV}$ (dotted). These coincide at small masses for a given reheating temperature, since $\delta<<m_{\chi_1}$. But they begin to diverge for larger masses, $m_{\chi_1}\gtrsim T_{RH}$, as the Boltzmann suppression leads to an exponential that is more sensitive to the difference of the two masses. \\
        We can see that the bounds from SN1987A are applicable on parts of these contours. 
        \item The total number of events at various xenon based DD experiments
        are shown in the color palette in fig.\,\ref{fig:ch4_EDM_res}. 
        The points shown correspond to a mass splitting of $\delta=1\,\text{keV}$. 
        We find that masses less than 12\,MeV and $4\times 10^{-6}\,{\rm GeV}^{-1}\lesssim d_\chi\lesssim 10^{-5}\,{\rm GeV}^{-1}$ are ruled out by XENONnT. These results are to be understood to be correct upto $\mathcal{O}(10\%)$ since we ignore astrophysical uncertainties (solar parameters and DM halo distribution) in probing orders of magnitude of DM masses.

\begin{figure}[ht!]
    \centering
\begin{subfigure}[b]{0.49\textwidth}
    \includegraphics[width=\textwidth]{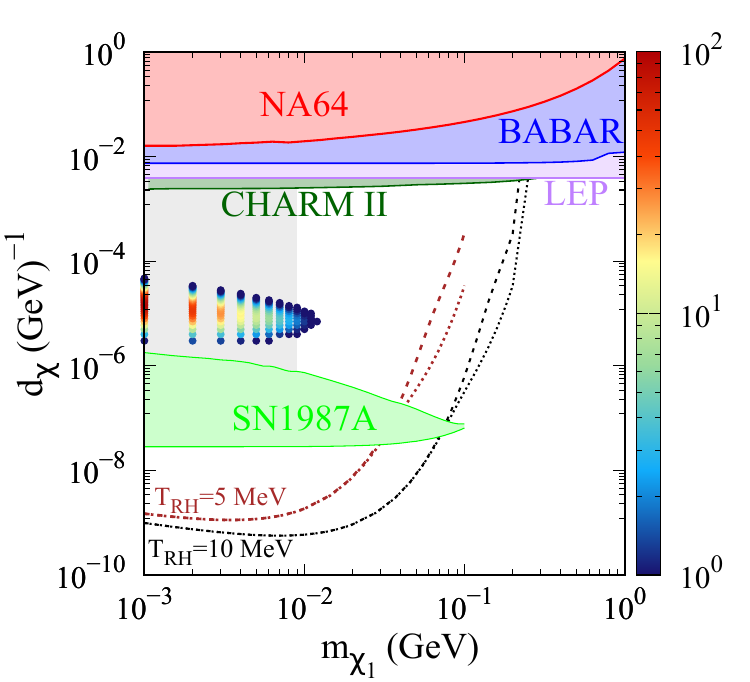}
    \caption{XENONnT}
\end{subfigure}
\begin{subfigure}[b]{0.49\textwidth}
    \includegraphics[width=\textwidth]{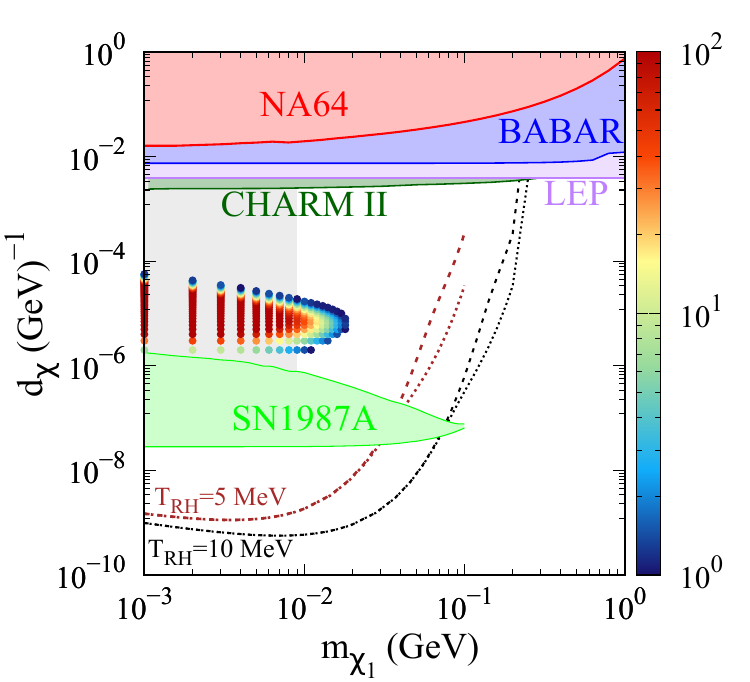}
    \caption{XENONnT projection}
\end{subfigure}
\begin{subfigure}[b]{0.49\textwidth}
    \includegraphics[width=\textwidth]{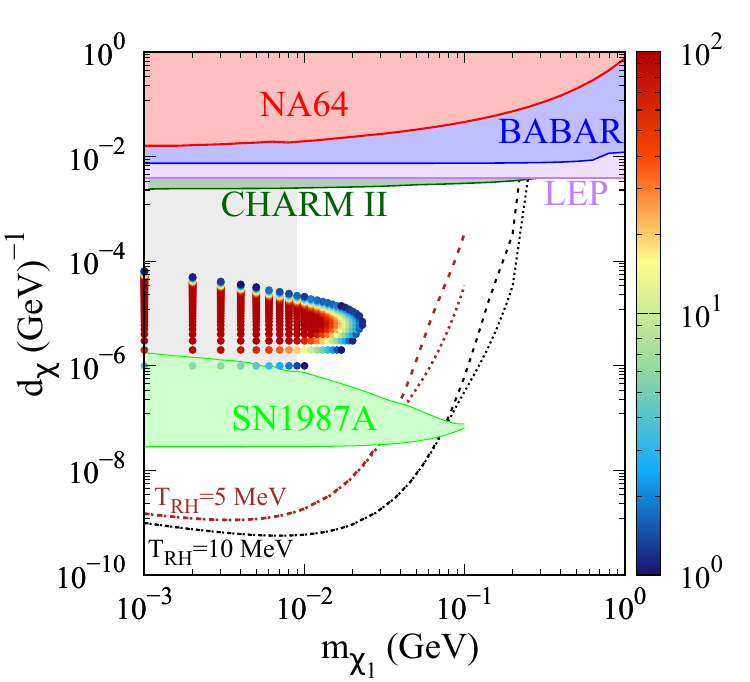}
    \caption{DARWIN}
\end{subfigure}
\caption{Constraints (shaded regions) on the transition electric dipole moment (EDM) from NA64 (red), BABAR (blue), LEP (purple), CHARM-II (dark green) and SN1987A (green). The gray shaded region is ruled out by $N_{\rm eff}$ constraints in standard cosmology\,\cite{Chang:2019xva}. Also shown are the contours that lead to observed relic density for $T_{RH}=$ 5\,MeV (brown) and 10\,MeV (black). The dashed and dotted lines for each correspond to mass splittings of $\delta=$ 1\,keV and 10\,keV, respectively. The points with color palette show the total number of events 
for mass splitting $\delta=1\,\text{keV}$ for various xenon experiments, as mentioned below each figure.}
\label{fig:ch4_EDM_res}
\end{figure}

\item We note that there is a competition between the decay rate of $\chi_2$ and its down-scattering cross section at DD experiments. This is because increasing $d_\chi$ gives larger scattering cross sections as seen from eq.\,\eqref{eq:sigmabare}, but also increases the decay width, $\Gamma_{\chi_2}\propto d_\chi^2$, so that the flux received on the Earth decreases. Therefore, we see from fig.\,\ref{fig:ch4_EDM_res} that starting from the smallest values of $d_\chi$, the rate initially increases with increasing $d_\chi$, maximizing at some value of $d_\chi$ and then falls quickly with further increase in $d_\chi$. 
\item This competition also leads to a maximum DM mass that can be probed at DD experiments. The XENONnT experiment currently probes masses $m_{\chi_1}\lesssim$12 MeV,  while in the future XENONnT can probe $m_{\chi_1}\lesssim 18\,$MeV and DARWIN can probe $m_{\chi_1}\lesssim23\,$MeV. 
We note that a large part of the points probed by XENONnT lie in the parameter space ruled out by $N_{\rm eff}$ bounds\,\cite{Chang:2019xva} for this model. These arise because of thermalization of the dark sector for large enough couplings,  assuming standard cosmology\footnote{These constraints might be evaded for non-standard cosmological evolution but we do not discuss them and choose to only show the reach of DD experiments for this parameter space as well.}.
\item Larger parts of the parameter space that aren't ruled out by $N_{\rm eff}$ bounds will be probed by future runs of XENONnT and DARWIN. Further, we show the event shapes for some benchmark points in figs.\,\ref{fig:ch4_rates_1}, \ref{fig:ch4_rates_2} and \ref{fig:ch4_rates_1.5del},  showing in red the signal+background rates, and the blue band corresponds to background rates with Poissonian uncertainties ($\pm\sqrt{N}$). 
We see that the signals may show up as an excess in the lowest recoil energy bins, and can also be distinguished from the background by the shape of the spectra. 

We can see the decrease in event rates when $\delta$ is increased from 1\,keV in fig.\,\ref{fig:ch4_rates_1} to 1.5\,kev in fig.\,\ref{fig:ch4_rates_1.5del}. 
This is because as $\delta$ increases, the decay width of $\chi_2$ increases, $\Gamma_{\chi_2}\propto\delta^3$. In addition, the up-scattering rates in the Sun get suppressed since the only electrons with enough energy to cause the up-scattering are those from the high velocity tail of MB distribution, given an average temperature of $T_\odot\simeq1.1$\,keV. Together, these prohibit the detection of large mass splittings ($\delta\gtrsim 1.2$\,keV at XENONnT current run, $\delta\gtrsim 1.7$\,keV at XENONnT future projection  and $\delta\gtrsim 1.8$\,keV at DARWIN) via electron scattering. 
\item The minimum $d_{\chi}$ values that can be probed by DD experiments are such that the parameter space that leads to the production of observed relic density by FI production, is not within the reach of the DD experiment. While smaller reheating temperatures could lead to an overlap between the two, its value is bounded from below by $T_{RH}>4$ MeV\,\cite{Hannestad:2004px}.
        \item We also show the complementary bounds from other experiments/ observations\,(NA64, BaBar, LEP and SN cooling) in fig.\,\ref{fig:ch4_EDM_res} for completeness. 
    \end{itemize}
\newpage
\begin{figure}[h!]
\centering
\begin{subfigure}[b]{\textwidth}
\begin{center}
        \includegraphics[width=0.45\textwidth]{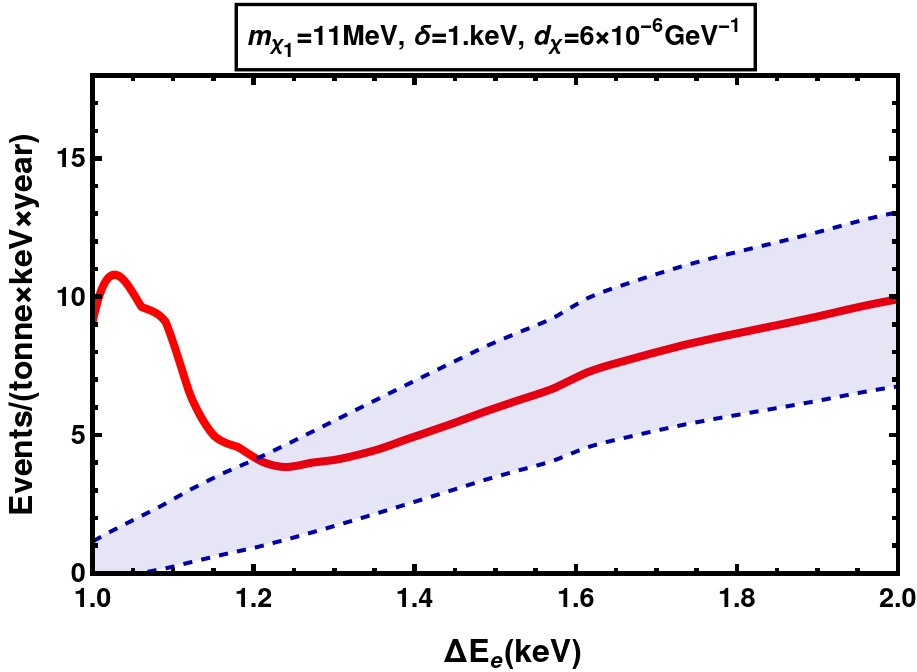}    
\end{center}
    \caption{ $m_{\chi_1}=11$\,MeV, $\delta=1\,$keV at XENONnT}        
\end{subfigure}
\begin{subfigure}[b]{0.47\textwidth}
    \includegraphics[width=\textwidth]{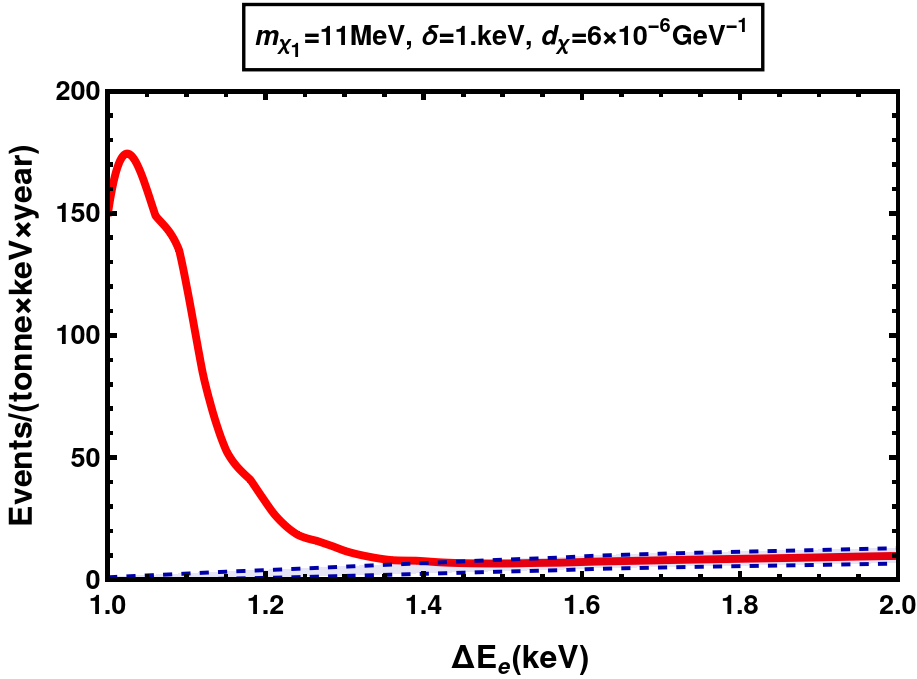}
    \caption{$m_{\chi_1}=11$\,MeV, $\delta=1\,$keV at XeNONnT {\small (proj.)} }
\end{subfigure}
\begin{subfigure}[b]{0.47\textwidth}
    \includegraphics[width=\textwidth]{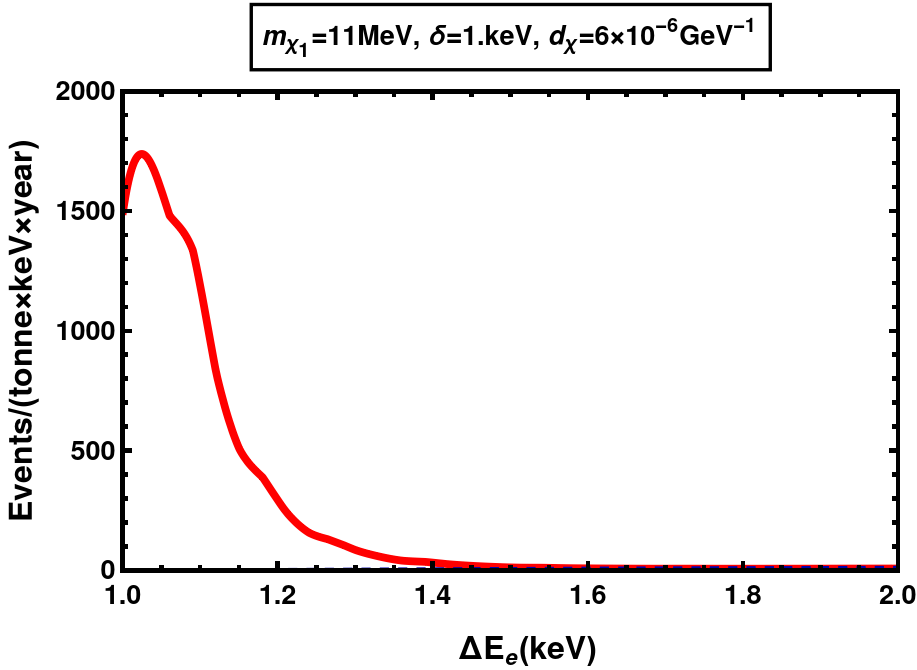}
    \caption{$m_{\chi_1}=11$\,MeV, $\delta=1\,$keV at DARWIN}
\end{subfigure}
\caption{Differential event rates for EDM DM  with $\delta=1.0$\,keV. Shown in blue are the background rates with the band representing Poissonian $(\pm\sqrt{N})$ uncertainties and in red are the signal+background rates.}
\label{fig:ch4_rates_1}
\end{figure}
\begin{figure}[h!]
\centering
\begin{subfigure}[b]{0.47\textwidth}
    \includegraphics[width=\textwidth]{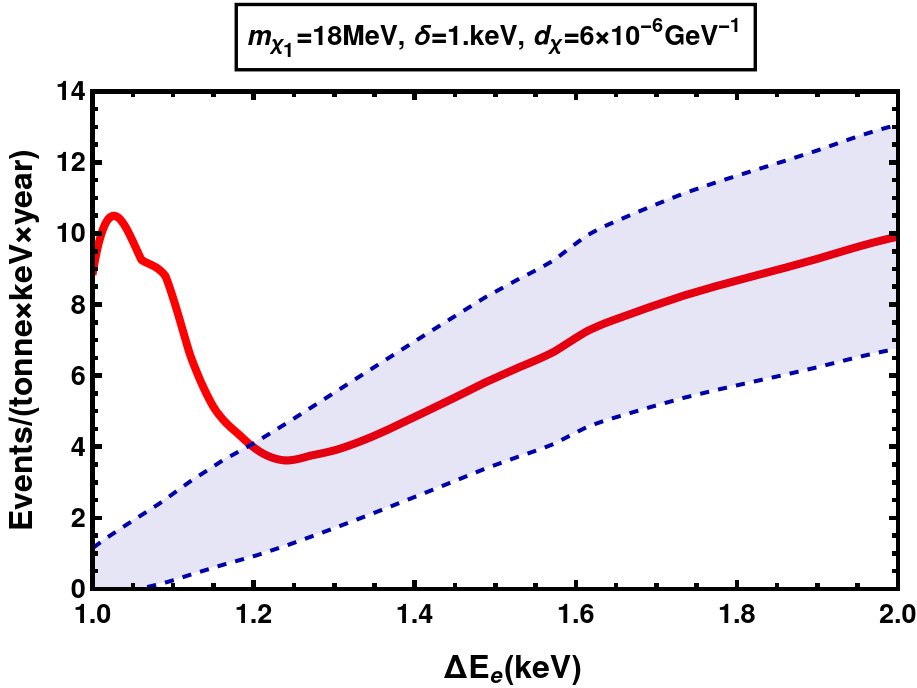}
    \caption{ $m_{\chi_1}=18$\,MeV, $\delta=1\,$keV at XENONnT {\small (proj.)}}
\end{subfigure}
\begin{subfigure}[b]{0.47\textwidth}
    \includegraphics[width=\textwidth]{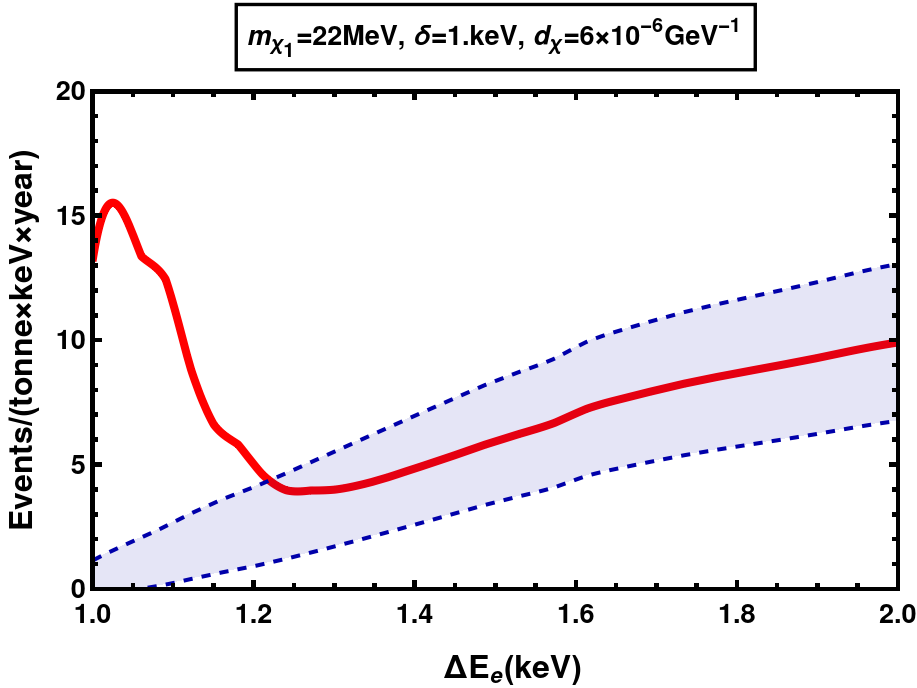}
    \caption{$m_{\chi_1}=22$\,MeV, $\delta=1\,$keV at DARWIN}
\end{subfigure}
\caption{Differential event rates for EDM DM  with $\delta=1.0\,$keV. Shown in blue are the background rates with the band representing Poissonian $(\pm\sqrt{N})$ uncertainties and in red are the signal+background rates.}
\label{fig:ch4_rates_2}
\end{figure}
\begin{figure}[h!]
\centering
\begin{subfigure}[b]{0.47\textwidth}
    \includegraphics[width=\textwidth]{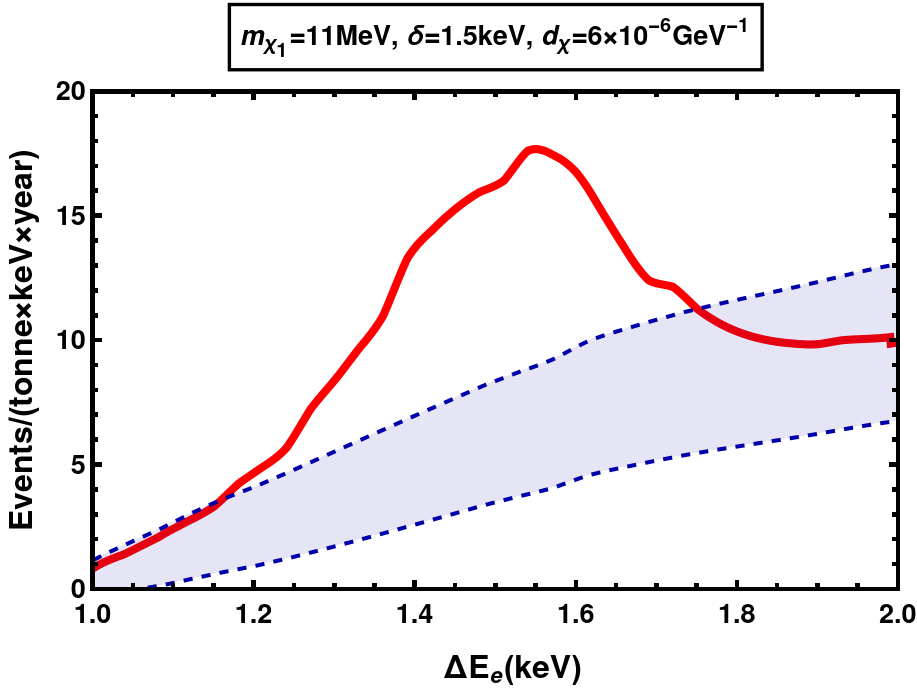}
    \caption{$m_{\chi_1}=11{\rm MeV}$,\,$\delta=1.5{\rm keV}$ at XENONnT {\small (proj.)}}
\end{subfigure}
\begin{subfigure}[b]{0.47\textwidth}
    \includegraphics[width=\textwidth]{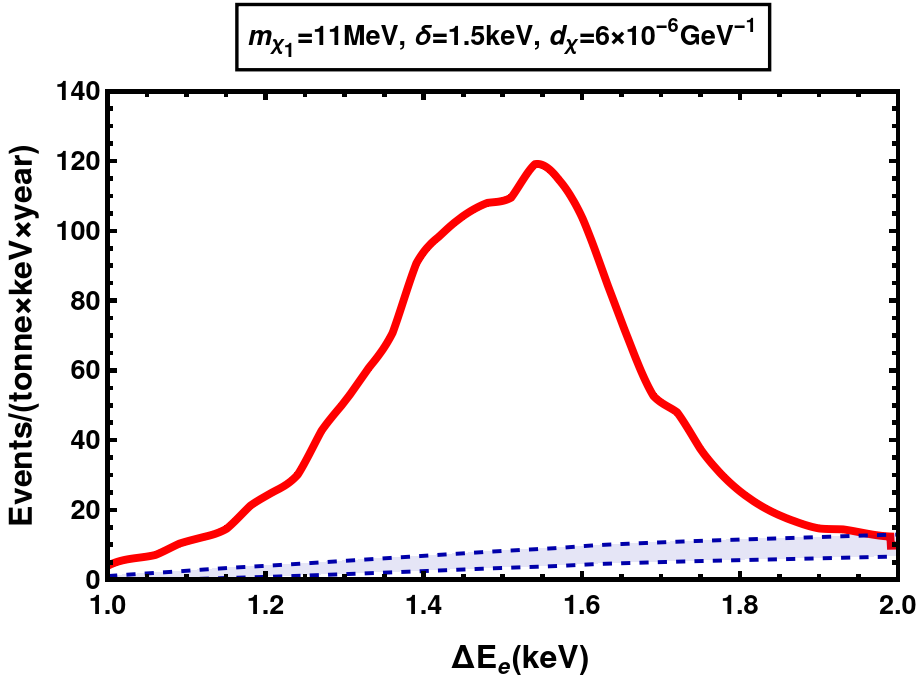}
    \caption{$m_{\chi_1}=11$\,MeV, $\delta=1.5\,$keV at DARWIN}
\end{subfigure}
\caption{Differential event rates for EDM DM with $\delta=1.5\,$keV. Shown in blue are the background rates with the band representing Poissonian $(\pm\sqrt{N})$ uncertainties and shown in red are the signal+background rates.}
\label{fig:ch4_rates_1.5del}
\end{figure}
\begin{figure}[h!]
\centering
\includegraphics[width=0.5\textwidth]{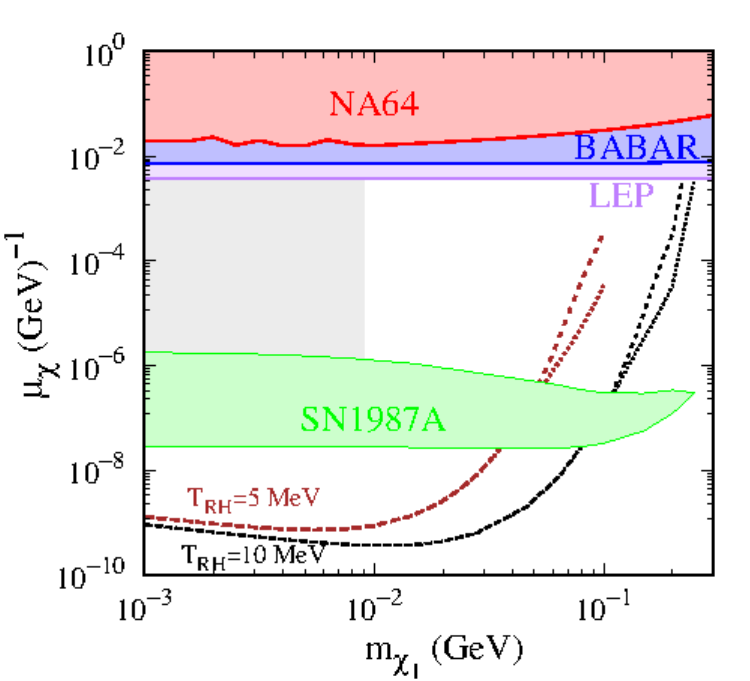}
\caption{Constraints (shaded regions) on the transition magnetic dipole moment (MDM) from NA64 (red), BABAR (blue), LEP (purple), CHARM-II (dark green) and SN1987A (green). The gray shaded region is ruled out by $N_{\rm eff}$ constraints in standard cosmology\,\cite{Chang:2019xva} for this model. These arise because of thermalization of the dark sector for large enough couplings,  assuming standard cosmology. Also shown are the contours that lead to observed relic density for $T_{RH}=$ 5\,MeV (brown) and 10\,MeV (black). The dashed and dotted lines for each correspond to mass splittings of $\delta=$ 1\,keV and 10\,keV, respectively.}
\label{fig:ch4_mdm_res}
\end{figure}
\newpage
\item Transition MDM interaction:\\
Here, we only discuss the features that are distinct from the EDM case, with all other features being the same.

\begin{itemize}
    \item We note from eqs.\,\eqref{eq:down_diff_cs} and \eqref{eq:sigmabare} that the MDM differential cross section $d\sigma/dE_R\propto \{v^2/E_R,1\}$ (corresponding to the two form factors $F_{DM}=1,1/q$, see eq.\,\eqref{eq:sigmabare}) which is suppressed with respect to the  corresponding EDM $d\sigma/dE_R\propto 1/v^2E_R$ by factors of $\{v^4,v^2E_R\}$. Here we use the fact that $q^2\simeq 2m_eE_R$ and the suppression comes from $v$ and $E_R$ being small. Therefore, the scattering rates for MDM interaction are highly suppressed and do not lead to significant event rates at the  xenon DD experiments, thus are not shown in fig.\,\ref{fig:ch4_mdm_res}. 
\end{itemize}
\end{enumerate}
\newpage
\section{Conclusions and Outlook}\label{sec:conclusions}
We have studied a model of inelastic DM interacting with SM via transition electric and magnetic dipole moments. We first address the production of DM by the FI mechanism taking into account both $2\rightarrow 2$ annihilation production and production from decay of plasmon. We find that both these processes are UV dominant with the rate of production (or $dY/dT$) maximised at the largest temperatures (near $T_{RH}$).  
We observe that the plasmon production is  insignificant for the parameter space we are interested in here, although it can become the dominant source of production for much larger reheating temperatures.

The heavier states are not stable on cosmological scales and their flux is produced by Solar up-scattering of the lighter, stable $\chi_1$  particles that are assumed to make up the entirety of DM. 
We study the constraints on this model from DD experiments where we can observe the down-scattering of the heavier mass state in electron recoil events. 
We find that DM with masses less than 12\,MeV and transition EDM $4\times 10^{-6}\,{\rm GeV}^{-1}\lesssim d_\chi\lesssim 10^{-5}\,{\rm GeV}^{-1}$ are ruled out by XENONnT\,\cite{XENONCollaboration:2022kmb}, already probing parameter space not ruled out by any other constraints. 

In addition, we find that future results from DD experiments, XENONnT and DARWIN, by virtue of larger exposures and lower backgrounds, can lead to the discovery of pseudo-Dirac DM with EDM interaction and mass splittings less than 2\,keV. Notably, this parameter space is not probed by any current experiment. The reach of DD experiments will further improve by lowering of detection thresholds (as suggested by the S2 only analysis from the XENON1T experiment\,\cite{XENON:2019gfn}).
We also show complementary constraints on the transition dipolar DM model from current fixed target experiments, $ e^+e^-$ colliders and information from SN cooling.
Projections from Belle-II  show that additional parameter space will be probed in the future\,\cite{Chu:2018qrm}. 

We thus study the most minimal model of inelastic DM with electric and magnetic dipolar couplings.
With a focus on  xenon based direct detection  experiments we find that future DD experiments have a great potential to discover this minimal model. Our work provides further motivation for an in-depth exploration of low-energy electron recoil events in  xenon based DD experiments.

\section{Acknowledgements}
We thank  Itay Bloch, Jae Hyeok Chang, Xiaoyong Chu,  Hyun Min Lee, Hongwan Liu,  Tarak Nath Maity, and Mukul Sholapurkar  for correspondence and helpful discussions. RL acknowledges financial support from the Infosys foundation (Bangalore), institute start-up funds, and
Department of Science and Technology (Govt.\,of India) for the grant SRG/2022/001125.
\appendix
\section{Plasmon Production}\label{app_plasmon}
The plasmon frequency is a good measure of the magnitude of medium effects and is given as\,\cite{Braaten:1993jw,Raffelt:1996wa}:
\begin{eqnarray}
\omega_p^2&=&\sum_{\psi\in SM}
\frac{4\alpha}{\pi}\int_0^\infty dp\, \frac{p^2}{E}\left(1-\frac{1}{3}v^2\right)(f_{\psi}+f_{\bar{\psi}})\\
&=&\sum_{\psi\in SM}\frac{4\alpha}{\pi}\int_0^{\infty} dp\, \frac{p^2}{\sqrt{p^2+m_\psi^2}}\left(1-\frac{1}{3}\frac{p^2}{p^2+m_\psi^2}\right)\left(\frac{1+\Theta(T-T_{\bar{\psi}})}{e^{\sqrt{p^2+m_\psi^2}/T}+1}\right),
\end{eqnarray}
where the sum is over contribution from SM fermions $\psi$. The velocity of SM particle $\psi$ is given by $v=p/E$ and the Fermi-Dirac distributions corresponding for SM fermions and anti-fermions are given by $f_\psi$ and $f_{\bar{\psi}}$,
respectively. To arrive at the second line, we assume that the chemical potentials are zero and that each antiparticle $\bar{\psi}$ stops contributing at temperature $T_{\bar{\psi}}$, which we take to be $\text{Max}[2m_{\bar{\psi}},\Lambda_{QCD}]$. 
The first mode frequency of the plasma is given by:
\begin{align}
\omega_1^2&=\sum_{\psi\in SM}
\frac{4\alpha}{\pi}\int_0^\infty dp\, \frac{p^2}{E}\left(\frac{5}{3}v^2-v^4\right)(f_{\psi}+f_{\bar{\psi}})\\
&=\sum_{\psi\in SM}\frac{4\alpha}{\pi}\int_0^{\infty} dp\, \frac{p^2}{\sqrt{p^2+m_e^2}}\left(\frac{5}{3}\frac{p^2}{p^2+m_e^2}-\frac{p^4}{(p^2+m_e^2)^2}\right)\left(\frac{1+\Theta(T-T_{\bar{\psi}})}{e^{\sqrt{p^2+m^2}/T}+1}\right),
\end{align}
using which we can define a velocity $v_*=\omega_1/\omega_p$. If we consider $\psi\in \{e\}$, then $v_*$ can be understood as the typical electron velocity.  
With these definitions, the general dispersion relations for the transverse and longitudinal polarizations are given by the following approximate expressions \cite{Braaten:1993jw} 
\begin{equation}\label{genwt}
\omega_T^2=|\vec{k}|^2+\omega_p^2\frac{3\omega_T^2}{2v_{\star}^2|\vec{k}|^2}\left(1-\frac{\omega_T^2-v_{\star}^2|\vec{k}|^2}{2\omega_T v_\star |\vec{k}|}\,\text{ln}\frac{\omega_T+v_{\star}|\vec{k}|}{\omega_T-v_{\star}|\vec{k}|}\right),\;\; 0 \leq |\vec{k}|<\infty,
\end{equation}
\begin{equation}\label{genwl}
\omega_L^2=\omega_p^2\frac{3\omega_L^2}{v_{\star}^2|\vec{k}|^2}\left(\frac{\omega_L}{2v_\star |\vec{k}|}\,\text{ln}\frac{\omega_L+v_{\star}|\vec{k}|}{\omega_L-v_{\star}|\vec{k}|}-1\right),\;\;0\leq |\vec{k}|\leq k_{max},
\end{equation}
that are correct to order $\alpha$. Here, $k_{max}$ is the maximum wavenumber upto which longitudinally polarized plasmons can be populated
\begin{equation}
k_{max}=\omega_p \left[\frac{3}{v_*^2}\left(\frac{1}{2v_*}\text{ln}\frac{1+v_*}{1-v_*}-1\right)\right]^{1/2}.
\end{equation} 
The in-medium couplings of the photon to
the SM particles are modified by vertex renormalization
constants $Z_{T,L}$ given by \cite{Raffelt:1996wa}
\begin{eqnarray}
Z_T(k)&=&\frac{2\omega_T^2(\omega_T^2-v_{\star}^2|\vec{k}|^2)}{3\omega_p^2\omega_T^2+(\omega_T^2+|\vec{k}|^2)(\omega_T^2-v_{\star}^2|\vec{k}|^2)-2\omega_T^2(\omega_T^2-|\vec{k}|^2)},\\
Z_L(k)&=&\frac{2(\omega_L^2-v_{\star}^2|\vec{k}|^2)}{3\omega_p^2-(\omega_L^2-v_{\star}^2|\vec{k}|^2)}\frac{\omega_L^2}{\omega_L^2-|\vec{k}|^2}.
\end{eqnarray}

Since the dispersion relations of transverse and longitudinal polarizations of the thermal photons are distinct, we separate the two polarizations in the follwoing calculation. The decay width of a plasmon with four momentum $k=(\omega,\vec{k})$ in the medium frame, and a definite polarization is \cite{Raffelt:1996wa}:
\begin{equation}
    \Gamma_{T,L}=\int \frac{d^3p_{\chi_2}}{(2\pi)^3\,(2E_{\chi_2})}\frac{d^3p_{\chi_1}}{(2\pi)^3\,(2E_{\chi_1})}(2\pi)^4\delta^4\left(k-p_{\chi_1}-p_{\chi_2}\right)\frac{1}{2\omega_{T,L}}\sum_{\text{spins}} |\mathcal{M}|_{\gamma^*\rightarrow \chi_1,\chi_2}^2
\end{equation}
where the squared amplitude, summed over incoming and outgoing spin states is
\begin{align}\label{plasmon_modm2}
\sum_{\text{spins}}|\mathcal{M}|_{\gamma^*\rightarrow \chi_1,\chi_2}^2=
    \begin{cases}
    4 d_\chi^2 Z_{T,L}\epsilon_\nu\epsilon^{*}_\sigma(m_{\chi_1}m_{\chi_2} k^2 g^{\nu\sigma}+k^2 g^{\nu\sigma}p_{\chi_1}.p_{\chi_2}  \\ \qquad-2g^{\nu\sigma}(k.p_{\chi_1})(k.p_{\chi_2})-k^2\left(p_{\chi_1}^\sigma p_{\chi_2}^\nu+p_{\chi_2}^\sigma p_{\chi_1}^\nu\right)),& \text{ for EDM,}\\
    4 \mu_\chi^2 Z_{T,L}\epsilon_\nu\epsilon^{*}_\sigma(-m_{\chi_1}m_{\chi_2} k^2 g^{\nu\sigma}+k^2 g^{\nu\sigma}p_{\chi_1}.p_{\chi_2}  \\ \qquad-2g^{\nu\sigma}(k.p_{\chi_1})(k.p_{\chi_2})-k^2\left(p_{\chi_1}^\sigma p_{\chi_2}^\nu+p_{\chi_2}^\sigma p_{\chi_1}^\nu\right)),& \text{ for MDM.}
    \end{cases}
\end{align}
The first term for each case in eq.\,\eqref{plasmon_modm2} is integrated in the rest frame of the plasmon as shown in eqs.\,(\ref{term0_inel}-\ref{term0_inel_simp}), giving:
\begin{equation}\label{term0_inel_simp_2}
\int \frac{d^3p_{\chi_2}}{(2\pi)^3\,2E_{\chi_2}}\frac{d^3p_{\chi_1}}{(2\pi)^3\,(2E_{\chi_1})}(2\pi)^4\delta^4\left(k-p_{\chi_1}-p_{\chi_2}\right)=\frac{1}{4\pi}\frac{|p_\chi^*|_0}{E^0_{\chi_1}+E^0_{\chi_2}}=\frac{1}{4\pi}C_1,
\end{equation}
with $C_1$ as given in eq.\,\eqref{C1_inel}. The integration for the last three terms are done using Lenard's formula\,\cite{Lenard:1953zz} updated for the massive, inelastic case 
\begin{equation}
\int \frac{d^3p_\chi}{(2\pi)^3\,2E_\chi}\frac{d^3p_{\bar{\chi}}}{(2\pi)^3\,(2E_{\bar{\chi}})}(2\pi)^4\delta^4\left(k-p_\chi-p_{\bar{\chi}}\right)p^\mu_{\chi}p^\nu_{\bar{\chi}}=\frac{1}{96\pi}\left(A\,k^2g^{\mu\nu}+2B\,k^\mu k^\nu\right),\label{Lenard1}
\end{equation}
as derived in appendix \ref{app4:lenard}.
The expression for $A$, $B$ are given in eqs.\,(\ref{lenard_inel_A}-\ref{lenard_inel_B}). Together these give the decay width as
\begin{eqnarray}
\Gamma_{T,L}=\begin{cases}
\frac{d_\chi^2Z_{T,L}}{24\pi \omega_{T,L}(k,T)}\left( B\,m_{T,L}(k,T)^4-12\,C_1m_{\chi_1}m_{\chi_2}m_{T,L}(k,T)^2\right),& \text{for EDM,} \\
\frac{\mu_\chi^2Z_{T,L}}{24\pi \omega_{T,L}(k,T)}\left(B \,m_{T,L}(k,T)^4+12\,C_1m_{\chi_1}m_{\chi_2}m_{T,L}(k,T)^2\right),& \text{for MDM.}
\end{cases}
\end{eqnarray}
We show in fig.\,\ref{fig:plasmon_IRUV} the difference in plasmon frequencies and relic density from plasmon decay, with only electrons and positrons modifying the dispersion relations, and that with the contribution from all SM fermions taken into account:
\begin{figure}[hb!]
\centering
\begin{subfigure}[b]{0.45\textwidth}
    \includegraphics[scale=0.45]{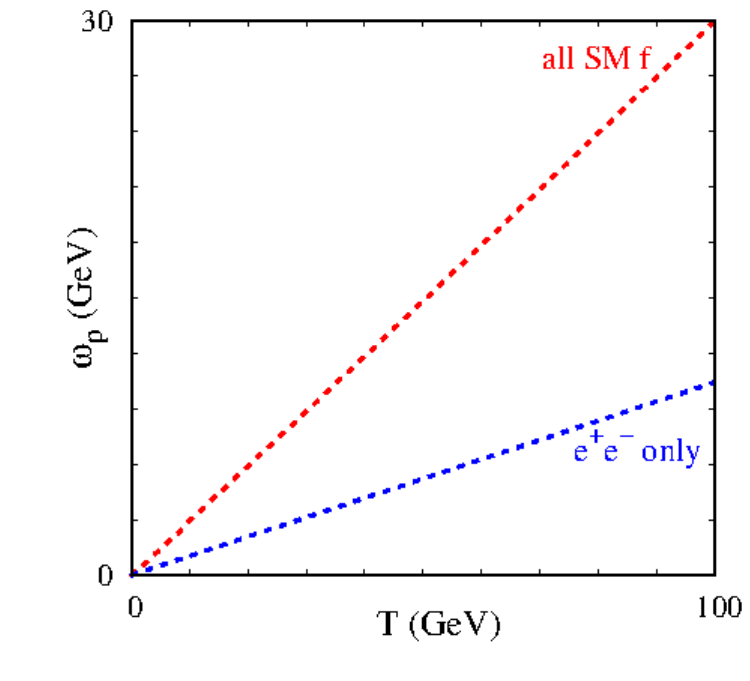}
    \label{fig:plasmon_IRUV_wp}
\end{subfigure}
\begin{subfigure}[b]{0.45\textwidth}
     \includegraphics[scale=0.45]{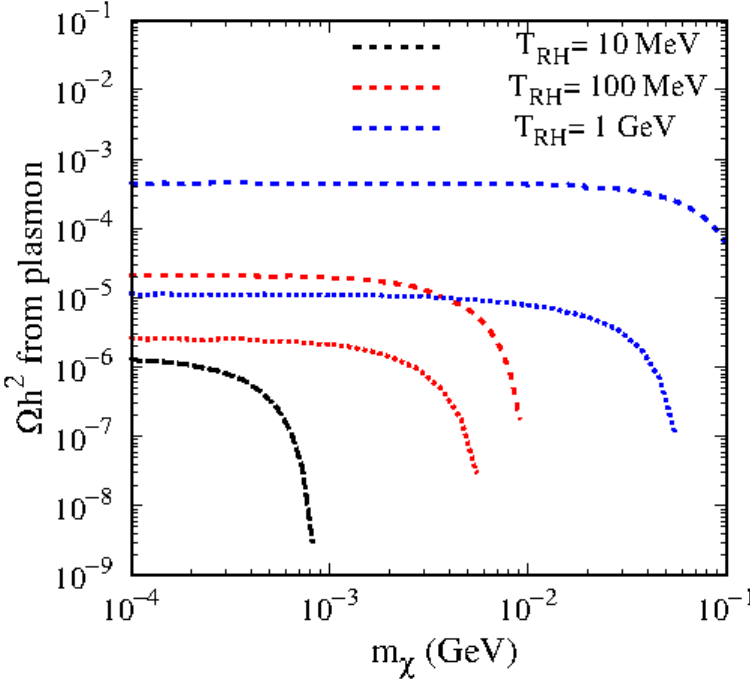}
    \label{fig:plasmon_IRUV_yield}
\end{subfigure}
\caption{To compare contributions from $e^+e^-$ only with that of all fermions: (a) plasma frequency for the two cases (b) DM relic density produced by considering electrons only (dotted lines) and all SM fermions (dashed lines) as a function of DM mass, for different reheating temperatures. For $T_{RH}=1\,$GeV, the two lines can be seen to coincide.}
\label{fig:plasmon_IRUV}
\end{figure}
\section{Lenard's Formula for Inelastic Case}\label{app4:lenard}
Lenard's formula for our case is \cite{Lenard:1953zz}:
\begin{equation}
\int \frac{d^3p_{\chi_1}}{(2\pi)^3\,(2E_{\chi_1})}\frac{d^3p_{\chi_2}}{(2\pi)^3\,(2E_{\chi_2})}(2\pi)^4\delta^4\left(k-p_{\chi_1}-p_{\chi_2}\right)p^\mu_{\chi_1}p^\nu_{\chi_2}=\frac{1}{96\pi}\left(A\,k^2g^{\mu\nu}+2B\,k^\mu k^\nu\right).\label{eq:inel_lenard}
\end{equation}
Multiplying both sides by $g^{\mu\nu}$ and contracting, we get
\begin{eqnarray}
k^2(4A+2B)&=& 96\pi\left(p_{\chi_1}.p_{\chi_2}\right) \int \frac{d^3p_{\chi_1}}{(2\pi)^3\,(2E_{\chi_1})}\frac{d^3p_{\chi_2}}{(2\pi)^3\,(2E_{\chi_2})}(2\pi)^4\delta^4\left(k-p_{\chi_1}-p_{\chi_2}\right)
\\
(4A+2B)&=&\frac{96\pi}{2}\left(1-\frac{m_{\chi_1}^2+m_{\chi_2}^2}{s}\right)\nonumber \\
&&\quad\times\int \frac{d^3p_{\chi_1}}{(2\pi)^3\,(2E_{\chi_1})}\frac{d^3p_{\chi_2}}{(2\pi)^3\,(2E_{\chi_2})}(2\pi)^4\delta^4\left(k-p_{\chi_1}-p_{\chi_2}\right)
\\
&=&12\;C_1\left(1-\frac{m_{\chi_1}^2+m_{\chi_2}^2}{s}\right)\label{lenard_inel_1}
\end{eqnarray}
where $k^2=s$. Here, the integration on RHS has been carried out in the rest frame of plasmon:
\begin{eqnarray}
&&\int \frac{d^3p_{\chi_2}}{(2\pi)^3(2E_{\chi_2})}\frac{d^3p_{\chi_1}}{(2\pi)^3(2E_{\chi_1})}(2\pi)^4\delta^4 \left(k-p_{\chi_1}- p_{\chi_2}\right)=\nonumber\\
&&\quad\quad\int \frac{d^3p_\chi^*}{(2\pi)^2\,4(E_{\chi_2}^*  E_{\chi_1}^*)} 
\times\,\delta^0\left(\omega^*-\sqrt{|p_\chi^*|^2+m_{\chi_2}^2}
-\sqrt{|p_\chi^*|^2+m_{\chi_1}^2}\right)\qquad\label{term0_inel}
\end{eqnarray}
where the $\vec{p}_{\chi_1}$ integral just sets $\vec{p}_{\chi_1}=\vec{p}_{\chi_2}$. Then we do the $\vec{p}_{\chi_2}$ integral in the rest frame of plasmon, defining all quantities in this frame with a $^*$ and redefining $p_{\chi_2}^*\equiv p_\chi^*$.

We simplify the delta function as
\begin{equation}
\delta^0\left(\omega^*-\sqrt{|p_\chi^*|^2+m_{\chi_2}^2}
-\sqrt{|p_\chi^*|^2+m_{\chi_1}^2}\right)=\frac{E^0_{\chi_1} E^0_{\chi_2} }{|p_\chi^*|_0(E^0_{\chi_1}+E^0_{\chi_2})}\delta^0\left(|p_\chi^*|-|p_\chi^*|_0\right),
\end{equation}
where,
\begin{eqnarray}
    |p_\chi^*|_0&=&\sqrt{\left(\frac{(\omega^*)^2+m_{\chi_2}^2-m_{\chi_1}^2}{2\omega^*}\right)^2-m_{\chi_2}^2}=\sqrt{\frac{\lambda\left((\omega^*)^2,m_{\chi_2}^2,m_{\chi_1}^2\right)}{4(\omega^*)^2}},\\
    {\rm and,}\qquad E^0_{\chi_i}&=&\sqrt{m_{\chi_i}^2+|p_\chi^*|_0^2}.
\end{eqnarray}
Here, $\lambda$ is the K\"{a}ll\'{e}n function defined as $\lambda(a,b,c)=a^2+b^2+c^2-2ab-2ac-2bc$.
Plugging this into eq.\,\eqref{term0_inel} we get
\begin{equation}\label{term0_inel_simp}
\int \frac{d^3p_{\chi_2}}{(2\pi)^3\,(2E_{\chi_2})}\frac{d^3p_{\chi_1}}{(2\pi)^3\,(2E_{\chi_1})}(2\pi)^4\delta^4\left(k-p_{\chi_1}-p_{\chi_2}\right)=\frac{1}{4\pi}\frac{|p_\chi^*|_0}{E^0_{\chi_1}+E^0_{\chi_2}}=\frac{1}{4\pi}C_1,
\end{equation}
\begin{equation}\label{C1_inel}
 \,\text{where,}\,\,\, C_1=\frac{\sqrt{\left(\frac{s+m_{\chi_2}^2-m_{\chi_1}^2}{2\sqrt{s}}\right)^2-m_{\chi_2}^2}}{\frac{s+m_{\chi_2}^2-m_{\chi_1}^2}{2\sqrt{s}}+\sqrt{\left(\frac{s+m_{\chi_2}^2-m_{\chi_1}^2}{2\sqrt{s}}\right)^2-m_{\chi_2}^2+m_{\chi_1}^2}}    
\end{equation}
Here, we have used $s=k^2=\omega_*^2\text{ (in CM frame}, \vec{k}_*=0 \implies k_*^2=\omega_*^2)$. \\Subsequently, multiplying both sides of eq.\,\eqref{eq:inel_lenard} by $k^\mu k^\nu$ and contracting, we get
\begin{eqnarray}
k^4(A+2B)=96\pi\;(k.p_{\chi_1})(k.p_{\chi_2})\int \frac{d^3p_{\chi_1}}{(2\pi)^3\,(2E_{\chi_1})}\frac{d^3p_{\chi_2}}{(2\pi)^3\,(2E_{\chi_2})}(2\pi)^4\delta^4\left(k-p_{\chi_1}-p_{\chi_2}\right).
\end{eqnarray}
Using $k=(p_{\chi_1}+p_{\chi_2}) $,
\[ (k-p_{\chi_1})^2=p_{\chi_2}^2\implies k^2+m_{\chi_1}^2-2\;k.p_{\chi_1}=m_{\chi_2}^2\implies k.p_{\chi_1}=\frac{k^2-m_{\chi_2}^2+m_{\chi_1}^2}{2},\]
\[ (k-p_{\chi_2})^2=p_{\chi_1}^2\implies k^2+m_{\chi_2}^2-2\;k.p_{\chi_2}=m_{\chi_1}^2\implies k.p_{\chi_2}=\frac{k^2-m_{\chi_1}^2+m_{\chi_2}^2}{2},\]
and substituting from eq.\,\eqref{term0_inel_simp}, we get:
\begin{equation}
(A+2B)=6C_1\left(1+\frac{m_{\chi_1}^2-m_{\chi_2}^2}{s}\right)\left(1+\frac{m_{\chi_2}^2-m_{\chi_1}^2}{s}\right)\label{lenard_inel_2}
\end{equation}
Putting together eqs.\,\eqref{lenard_inel_1} and \eqref{lenard_inel_2}, we get
\begin{equation}\label{lenard_inel_A}
A=2C_1\left(1+\left(\frac{m_{\chi_2}^2-m_{\chi_1}^2}{s}\right)^2-2\left(\frac{m_{\chi_1}^2+m_{\chi_2}^2}{s}\right)\right)
\end{equation}
\begin{equation}\label{lenard_inel_B}
B=2C_1\left(1+\left(\frac{m_{\chi_2}^2+m_{\chi_1}^2}{s}\right)-2\left(\frac{m_{\chi_2}^2-m_{\chi_1}^2}{s}\right)^2\right)
\end{equation}
for the inelastic Lenard's formula, eq.\eqref{eq:inel_lenard}, with $C_1$ given in eq.\,\eqref{C1_inel}

\section{Fixed Target: NA64 Events}\label{app4:NA64}
We review the discussion in ref.\,\cite{Chu:2018qrm} and begin by noting that in full generality, a 4-body phase space has 12 degrees of freedom. In particular, though, there are redundancies from  invariance in rotation around the beam line, and from imposition of energy-momentum conservation, leaving us with 7 independent degrees of freedom. With this knowledge, the 4-body phase space can be written as
\begin{equation}
    \frac{d\Phi_4}{ds_{3\chi_1\chi_2}dq_2^2}=\frac{|J|}{16(2\pi)^6}\frac{ds_{\chi_1\chi_2}}{s_{\chi_1\chi_2}}dq_1^2\frac{\lambda^{1/2}\left(s_{\chi_1\chi_2},m_{\chi_1}^2,m_{\chi_2}^2\right)}{\lambda^{1/2}\left(s_{3\chi_1\chi_2},m_N^2,q_2^2\right)}du_{2_q}\Bigg|\frac{\partial\phi_3^{R3\chi_1\chi_2}}{\partial u_{2_q}}\Bigg|\frac{d\Omega_\chi^{R\chi_1\chi_2}}{4\pi},\label{app4:phi4}
\end{equation}
where the kinematic quantities are,
\begin{align}
    s_{3\chi_1\chi_2}&=(p_3+p_{\chi_1}+p_{\chi_2})^2,\nonumber\\
    s_{\chi_1\chi_2}&=q^2=(p_{\chi_1}+p_{\chi_2})^2,\nonumber\\
    u_{2_q}&=p_2.q \nonumber
\end{align} 
Here, $q_2=p_2-p_4$, $ q_1=p_1-p_3$ and the remaining momenta are as shown in fig.\,\ref{fig:NA64}.
The azimuthal angle of $p_3$ in the frame where $\vec{p}_3+\vec{p}_{\chi_1}+\vec{p}_{\chi_2}=0$ is given by $\phi_3^{R3\chi_1\chi_2}$  and the  solid angle between the DM particles is $\Omega_\chi^{R\chi_1\chi_2}$. The K\"{a}ll\'{e}n function denoted by $\lambda$ is defined as
\[\lambda(a,b,c)=a^2+b^2+c^2-2ab-2ac-2bc.\]
The Jacobian of the transformation from $E_4,{\rm cos}\theta_4$ to $s_{3\chi_1\chi_2},q_2^2$, required to connect between eq.\,\eqref{eq:NA64_events} and eq.\,\eqref{app4:phi4}, is given by
\begin{equation}
    \frac{\partial(E_4,\textrm{cos}\,\theta_4)}{\partial(s_{3\chi_1\chi_2},q_2^2)}=\frac{|J|}{|\vec{p}_4|}\equiv\frac{\lambda^{-1/2}(s,m_N^2,m_e^2)}{2|\vec{p}_4|}
\end{equation}
The double differential cross section for the production corresponding to fig.\,\ref{fig:NA64} is
\begin{equation}
    \frac{d\sigma_{prod}}{dE_4d\textrm{cos}\,\theta_4}=\frac{|\vec{p}_4|}{4E_2m_N|\vec{v}_2|}\int\frac{d\Phi_4}{ds_{3\chi_1\chi_2}dq_2^2}\frac{1}{|J|}|\mathcal{M}|^2,\label{app4:NA64_diffsigma}
\end{equation}
where $E_2=E_0$ is the incoming electron (beam) energy with its velocity given as $|\vec{v}_2|=\sqrt{1-(m_e/E_0)^2}$.

The squared amplitude for the process shown in fig.\,\ref{fig:NA64} is 
\begin{equation}
    |\mathcal{M}|^2=(4\pi\alpha)^3\frac{2}{q^4q_1^4}L^{\rho\sigma,\mu\nu}\chi_{\mu\nu}(q)W_{\rho\sigma}(-q_1)\Bigg|_{s_X=m_N^2},
\end{equation}
with $\chi_{\mu\nu}$ the DM emission piece of the amplitude,
\begin{equation}
    \chi_{\mu\nu}=\textrm{Tr}\left[(\slashed{p}_{\chi_1}+m_{\chi_1})\Gamma_\mu(q)(\slashed{p}_{\chi_2}-m_{\chi_2})\bar{\Gamma}_\nu(q)\right].
\end{equation}
Here, the interaction operators are given by $\Gamma^\mu(q)=-d_\chi\sigma^{\mu\nu}q_\nu\gamma^5$ for EDM and $\Gamma^\mu(q)=i\mu_\chi\sigma^{\mu\nu}q_\nu$ for MDM.
The term corresponding to electron scattering, with averaging over initial and sum over final spins is 
\begin{align}
    L^{\rho\sigma,\mu\nu}&=\frac{L^{\rho\sigma,\mu\nu}_a}{[(p_4+q)^2-m_e^2]^2}+\frac{L^{\rho\sigma,\mu\nu}_b}{[(p_2-q)^2-m_e^2]^2}+\frac{2L^{\rho\sigma,\mu\nu}_{ab}}{[(p_4+q)^2-m_e^2][(p_2-q)^2-m_e^2]}\\
    \textrm{with, \,} L^{\rho\sigma,\mu\nu}_a&=\frac{1}{2}\textrm{Tr}\left[(\slashed{p}_4+m_e)\gamma^\mu(\slashed{p}_4+\slashed{q}+m_e)\gamma^\rho(\slashed{p}_2+m_e)\gamma^\sigma(\slashed{p}_4+\slashed{q}+m_e)\gamma^\nu\right]\nonumber\\
    L^{\rho\sigma,\mu\nu}_b&=\frac{1}{2}\textrm{Tr}\left[(\slashed{p}_2+m_e)\gamma^\nu(\slashed{p}_2-\slashed{q}+m_e)\gamma^\sigma(\slashed{p}_4+m_e)\gamma^\rho(\slashed{p}_2-\slashed{q}+m_e)\gamma^\mu\right]\nonumber\\
    L^{\rho\sigma,\mu\nu}_{ab}&=\frac{1}{2}\textrm{Tr}\left[(\slashed{p}_4+m_e)\gamma^\mu(\slashed{p}_4+\slashed{q}+m_e)\gamma^\rho(\slashed{p}_2+m_e)\gamma^\sigma(\slashed{p}_2-\slashed{q}+m_e)\gamma^\nu\right]\nonumber.
\end{align}
The hadronic tensor describing the response of the nuclear target is
\begin{equation}
W^{\rho\sigma}\simeq\left(p_1^\rho-\frac{p_1.q_1}{q_1^2}q_1^\rho\right)\left(p_1^\sigma-\frac{p_1.q_1}{q_1^2}q_1^\sigma\right)\frac{W(q_1^2,s_X)}{m_N^2}.
\end{equation}
Assuming Pb to be a scalar target gives
\begin{align*}
W_1(q^2)&=0, \\
W_2(q^2)&=4m_N^2F_E^2(q_1^2)\delta(s_X-m_N^2)/2,
\end{align*}
with, $F_E(t)=Z\frac{a^2(Z)t}{1+a^2(Z)t}\frac{1}{1+t/d(A)}$, $t=-q_1^2$, $a(Z)=111\frac{Z^{1/3}}{m_e}$ and $d(A)=0.164$\,GeV$^2A^{-2/3}$.
The mass number and atomic number of the target nucleus  are given by $A$ and $Z$, respectively. We have neglected the magnetic form factor for $Z>>1$.

The integration boundaries for eq.\,\eqref{app4:NA64_diffsigma} are:
\begin{align*}
    &(m_N+m_{\chi_1}+m_{\chi_2})^2\leq s_{3\chi_1\chi_2}\leq(\sqrt{s}-m_e)^2, \\
    &(m_{\chi_1}+m_{\chi_2})^2\leq s_{\chi_1\chi_2}\leq (\sqrt{s_{3\chi_1\chi_2}}-m_N)^2,\\
    [q_1^2]^\pm&=2m_N^2-\frac{(s_{3\chi_1\chi_2}+m_N^2-q_2^2)(s_{3\chi_1\chi_2}+m_N^2-s_{\chi_1\chi_2})}{2s_{3\chi_1\chi_2}}\\
    &\mp\frac{\lambda^{1/2}(s_{3\chi_1\chi_2},m_N^2,q_2^2)\lambda^{1/2}(s_{3\chi_1\chi_2},m_N^2,s_{\chi_1\chi_2})}{2s_{3\chi_1\chi_2}},\\
    [q_2^2]^\pm&=2m_e^2-\frac{(s+m_e^2-m_N^2)(s+m_e^2-s_{3\chi_1\chi_2})}{2s}\mp \frac{\lambda^{1/2}(s,m_e^2,m_N^2)\lambda^{1/2}(s,m_e^2,s_{3\chi_1\chi_2})}{2s}.
\end{align*}
And the angular variable $u_{2_q}$ is given by
\begin{align*}
u_{2_q}=&\frac{(p_1.p_2)G_2(p_1,q_2;q_2,q)}{-\Delta_2(p_1,q_2)}-\frac{(q_2.p_2)G_2(p_1,q_2;p_1,q)}{-\Delta_2(p_1,q_2)}\\
&+\frac{\sqrt{\Delta_3(p_1,q_2,p_2)\Delta_3(p_1,q_2,q)}}{-\Delta_2(p_1,q_2)}\textrm{cos}\,\phi_3^{R3\chi_1\chi_2}
\end{align*}
where $G_n$ is the Gram determinant of dimension n and $\Delta_n$ is the Cayley determinant (symmetric Gram determinant)\,\cite{Han:2005mu,Byckling:1971vca}.\\
And finally, since the NA64 experiment isn't sensitive to the angular distribution of the outgoing DM particles, we can integrate over the solid angle between the DM particles, $\Omega_\chi^{R\chi_1\chi_2}$ in eq.\,\eqref{app4:phi4}. 

\section{Derivation of the electron recoil rate formula for upscattered DM}
We follow the discussion in Appendix A of Essig et al \cite{Essig:2015cda} to re-derive the electron scattering rate for a general DM flux.  

The integrated flux changes as follows in going from the Standard Halo Model DM distribution $\left(g_{\chi}(v)\right)$ to a generic incoming DM flux per unit energy $\left(d\Phi/dK_{\chi_2}\right)$
\begin{equation}
    \int dv\, g_{\chi}(v)\frac{\rho_\chi}{m_\chi} v\rightarrow \int dK_{\chi_2}\frac{d\Phi}{dK_{\chi_2}}\label{eq:app_gal_upscatter}
\end{equation}
Then, starting from eq.\,A12 of \cite{Essig:2015cda}, we rewrite the cross section for a $\chi_2$ to excite an electron from level 1 to level 2 of an atom as
\begin{eqnarray}
\sigma_{1\rightarrow 2}=\frac{\bar{\sigma}_e}{4\mu_{\chi_2,e}^2v}\int \frac{d^3q}{4\pi}\delta\left(\Delta E_{1\rightarrow 2}+\frac{q^2}{2m_{\chi_2}}-q\,v\,{\rm cos}\theta_{qv}\right)|F_{DM}(q)|^2|f_{1\rightarrow 2}(q)|^2\bigg|_{v=\sqrt{2K_{\chi_2}/m_{\chi_2}}}.\label{eq:app_e_exc}
\end{eqnarray}
The rate of excitation events, for a given transition and given target electrons, is found by multiplying eq.\,\eqref{eq:app_e_exc} by the incoming $\chi_2$ flux. Using eq.\,\eqref{eq:app_gal_upscatter} we get this to be
\begin{eqnarray}
R_{1\rightarrow 2}&=&\int dK_{\chi_2}\frac{d\Phi}{dK_{\chi_2}}\sigma_{1\rightarrow 2}\,\\
&=&\int dK_{\chi_2}\frac{d\Phi}{dK_{\chi_2}} 
\frac{\bar{\sigma}_e}{4\mu_{\chi_2,e}^2} \sqrt{\frac{m_{\chi_2}}{2K_{\chi_2}}}
\int \frac{d^3q}{4\pi} \delta \left(\Delta E_{1\rightarrow 2}+\frac{q^2}{2m_{\chi_2}}-q\sqrt{\frac{2K_{\chi_2}}{m_{\chi_2}}}{\rm cos}\theta_{qv}\right) \nonumber\\
&&\qquad\times|F_{DM}(q)|^2|f_{1\rightarrow 2}(q)|^2.
\end{eqnarray}
The rates for ionization of electrons bound in isolated atoms can be calculated with the simplifying assumptions of a spherically symmetric atomic potential and filled shells. The ionized electron can be treated as being in one of a continuum of positive-energy bound states, approximated to free particle states at asymptotically large radii. The ionization rate for an atom is found by taking eq.\,\eqref{eq:app_e_exc}, summing over occupied electron shells, and integrating over all possible final states. For ionization, with the final states being a continuum, the phase space is\,\cite{Essig:2015cda}
\begin{equation}
    \textrm{ionized electron phase space=}\sum_{l'm'}\int \frac{k'^2dk'}{(2\pi)^3}=\frac{1}{2}\sum_{l'm'}\int \frac{k'^3d{\rm ln}\,E_R}{(2\pi)^3}.
\end{equation}
Here, $l',m'$ are the angular quantum numbers of the ionized electron final state and $k'$ is its momentum at asymptotically large distances from the nucleus, with energy $E_R=k'^2/2m_e$. Plugging this in, the ionization rate is given as
\begin{eqnarray}
R_{\rm ion}&=&\sum_{\textrm{occ. states}}\sum_{l'm'}\int dK_{\chi_2}\frac{k'^3d{\rm ln}E_R }{2(2\pi)^3}\frac{d\Phi}{dK_{\chi_2}} \frac{\bar{\sigma}_e}{4\mu_{\chi_2,e}^2} \sqrt{\frac{m_{\chi_2}}{2K_{\chi_2}}}\nonumber\\
& &\qquad\times\int\frac{d^3q}{4\pi} \delta \left(\Delta E_{1\rightarrow 2}+\frac{q^2}{2m_{\chi_2}}-q\sqrt{\frac{2K_{\chi_2}}{m_{\chi_2}}}{\rm cos}\theta_{qK_{\chi_2}}\right)|F_{DM}(q)|^2|f_{i\rightarrow k'l'm'}(q)|^2\, ,\\
&=&\int dK_{\chi_2}d{\rm ln}E_R\frac{d\Phi}{dK_{\chi_2}}\frac{\bar{\sigma}_e}{16\mu_{\chi_2,e}^2}\sqrt{\frac{m_{\chi_2}}{2K_{\chi_2}}}\nonumber\\
& &\qquad\times\int\frac{d^3q}{4\pi} \delta \left(\Delta E_{1\rightarrow 2}+\frac{q^2}{2m_{\chi_2}}-q\sqrt{\frac{2K_{\chi_2}}{m_{\chi_2}}}{\rm cos}\theta_{qK_{\chi_2}}\right)|F_{DM}(q)|^2|f_{\rm ion}(k',q)|^2\,,
\end{eqnarray}
with 
\begin{equation}
    |f_{\rm ion}(k',q)|^2\equiv \frac{2k'^3}{(2\pi)^3}\sum_{\textrm{occ. states}}\sum_{l'm'}|f_{i\rightarrow k'l'm'}(q)|^2.
\end{equation}
We can write the electron recoil energy spectrum per detector mass per unit time as
\begin{equation}
    \frac{dR_{\rm ion}}{d\Delta E_e}=n_T\frac{\bar{\sigma}_e}{64\mu_{\chi_2,e}^2}\sum_{n,l}\frac{1}{\Delta E_e-E_{nl}}\int dK_{\chi_2}\frac{d\Phi}{dK_{\chi_2}}\frac{m_{\chi_2}}{K_{\chi_2}}\int dq\,q|F_{DM}(q)|^2|f_{\rm ion}(k',q)|^2  ,
\end{equation}
where the total energy transferred to the electron is $\Delta E_e=E_R+E_{nl}$ and $n_T$ is the number of targets per tonne.
To explicitly show the order of integration, including the factor for depletion in numbers  due to decay of $\chi_2$ in travelling from the Sun to the Earth, and the energy dependent detector efficiency ($\epsilon(\Delta E_e)$), we get:
\begin{eqnarray}
    R_{\rm ion}&=&n_T  \int d\Delta E_e \,\epsilon(\Delta E_e)
    \sum_{n,l}
    \frac{1}{\Delta E_e-E_{nl}}
    \frac{\bar{\sigma}_e}{64\mu_{\chi_2,e}^2}
    \int dK_{\chi_2}\Theta(\Delta E_e^{\rm max}(K_{\chi_2})-\Delta E_e)\frac{d\Phi}{dK_{\chi_2}}\frac{m_{\chi_2}}{K_{\chi_2}}
    e^{-t(K_{\chi_2})\times \Gamma_{\chi_2}}\nonumber\\
    & &\qquad\qquad\times \int_{q^-(K_{\chi_2},\Delta E_e,\delta,m_{\chi_2})}^{q^+(K_{\chi_2},\Delta E_e,\delta,m_{\chi_2})} dq\,q|F_{DM}(q)|^2|f_{nl\rightarrow \Delta E_e-E_{nl}}(q)|^2  .
\end{eqnarray}

\section{Effect of accounting for detector resolution in recoil spectra}
\begin{figure}[h!]
\centering
\begin{subfigure}[b]{0.47\textwidth}
    \includegraphics[width=\textwidth]{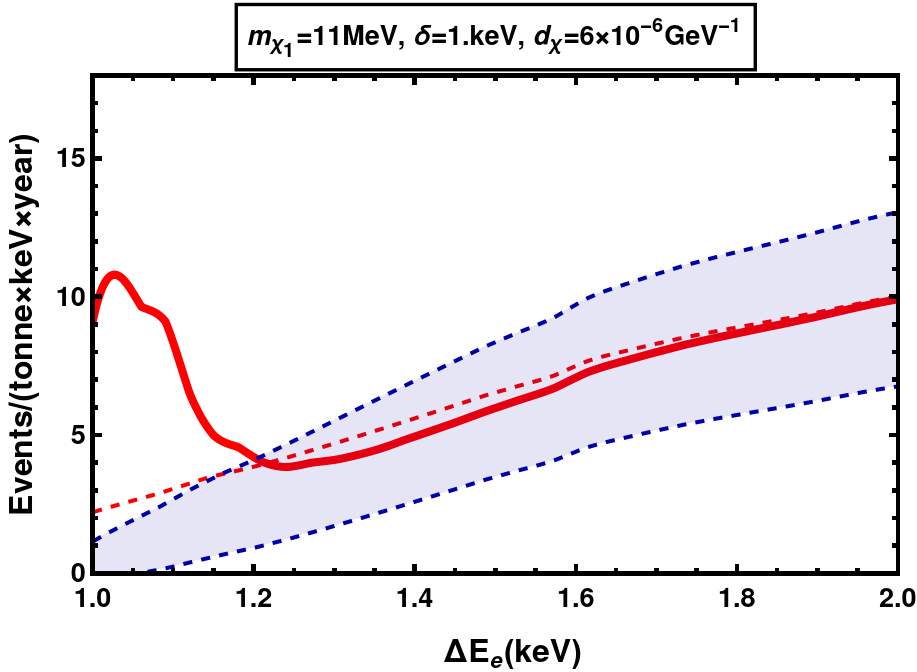} 
    \caption{$m_{\chi_1}=11$\,MeV, $\delta=1\,$keV at XENONnT}
\end{subfigure}\\

\begin{subfigure}[b]{0.47\textwidth}
    \includegraphics[width=\textwidth]{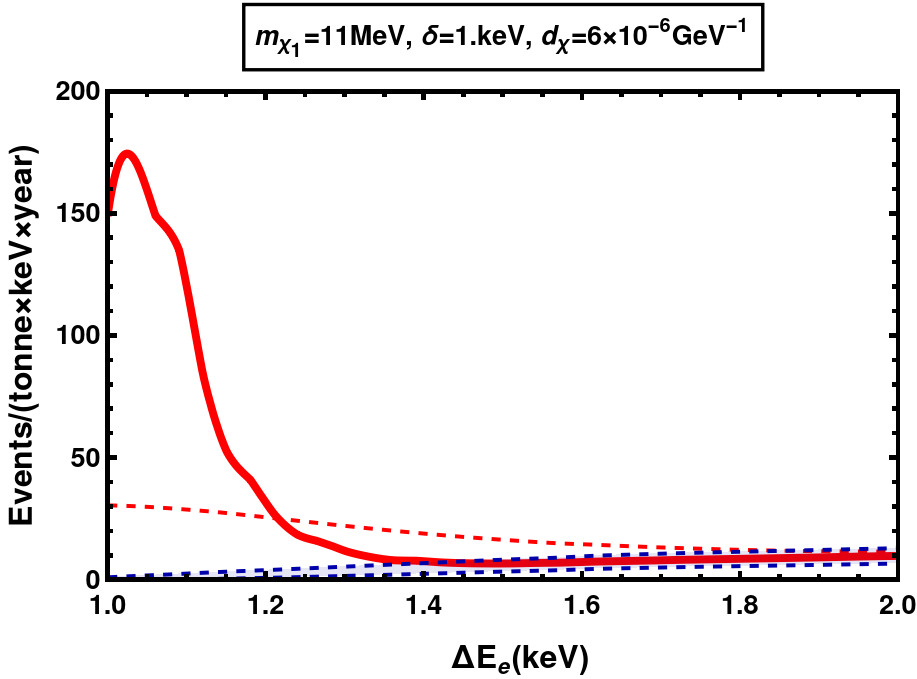}   
    \caption{$m_{\chi_1}=11$\,MeV, $\delta=1\,$keV at XeNONnT {\small (proj.)}}
\end{subfigure}
\begin{subfigure}[b]{0.47\textwidth}
    \includegraphics[width=\textwidth]{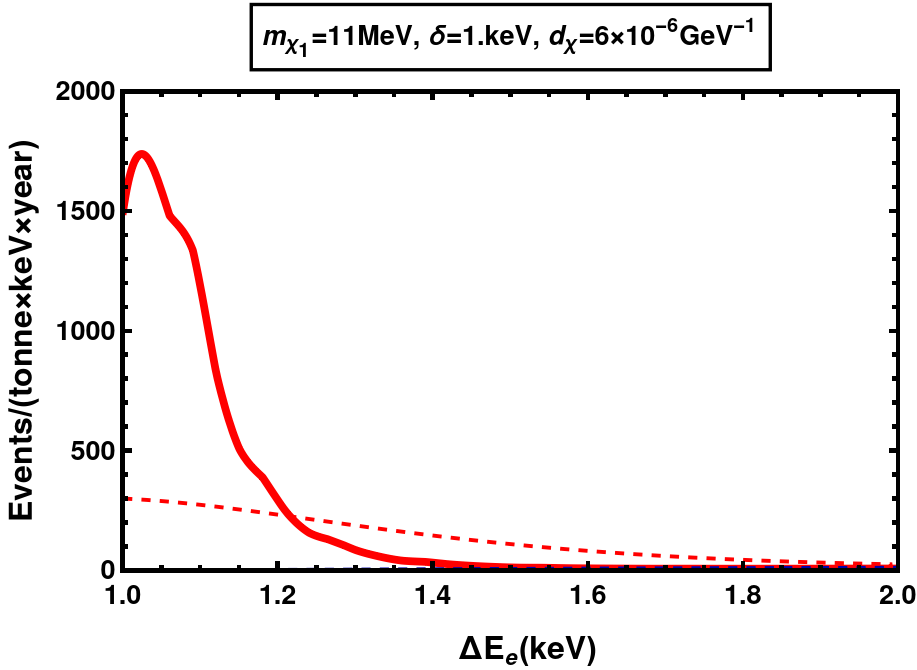}    
    \caption{$m_{\chi_1}=11$\,MeV, $\delta=1\,$keV at DARWIN}
\end{subfigure}

\begin{subfigure}[b]{0.47\textwidth}
    \includegraphics[width=\textwidth]{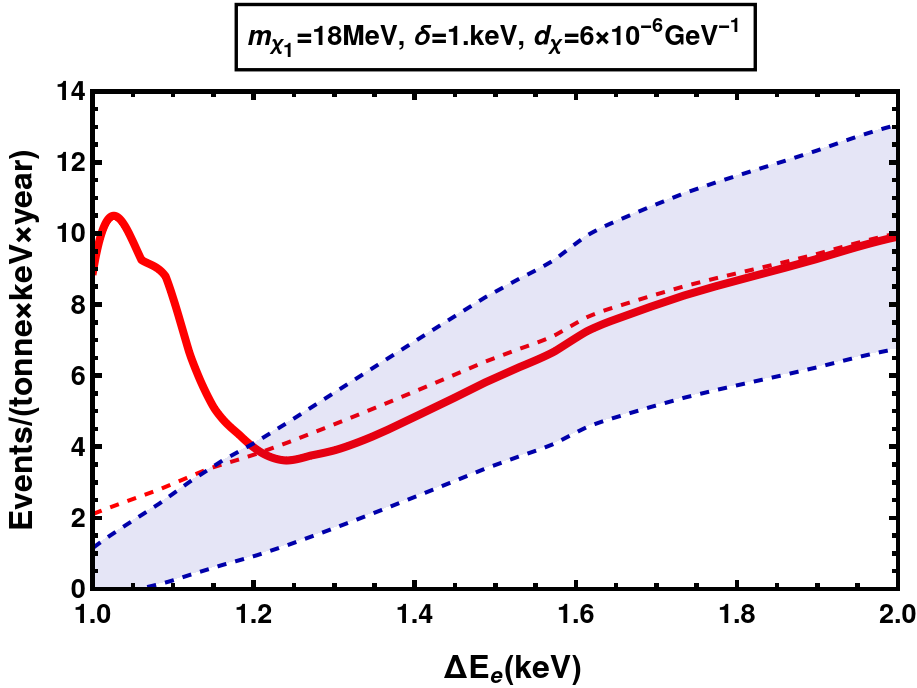}   
    \caption{$m_{\chi_1}=18$\,MeV, $\delta=1\,$keV at XENONnT {\small (proj.)}}
\end{subfigure}
\begin{subfigure}[b]{0.47\textwidth}
    \includegraphics[width=\textwidth]{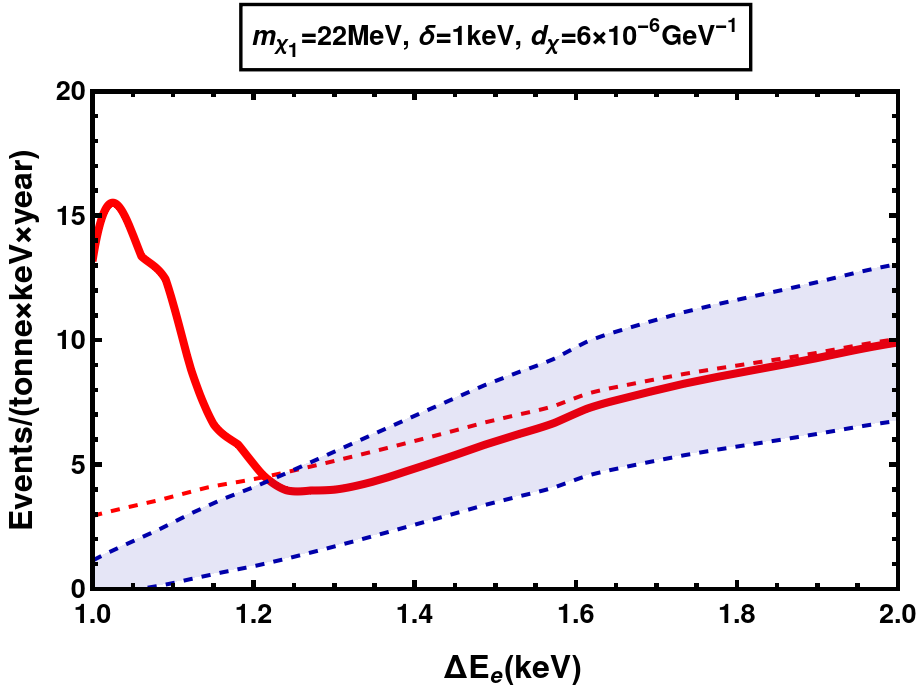} 
    \caption{$m_{\chi_1}=22$\,MeV, $\delta=1\,$keV at DARWIN}
\end{subfigure}
\caption{Differential event rates for EDM DM. Shown in blue are the background rates with the band representing Poissonian $(\pm\sqrt{N})$ uncertainties. The dashed (solid) red lines show the signal+background rates with (without) smearing from detector resolution, using eq.\,\eqref{eq:e_rec_spec_det} (eq.\,\eqref{eq:e_rec_spec}).}
\label{fig:smeared_spec1}
\end{figure}

\begin{figure}[h!]
\begin{subfigure}[b]{0.47\textwidth}
    \includegraphics[width=\textwidth]{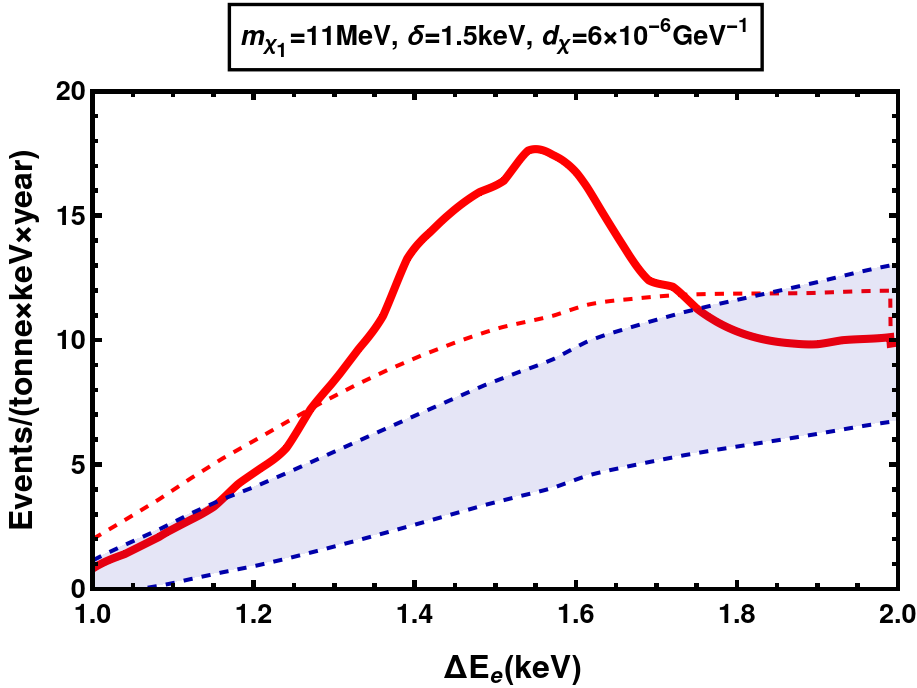}
    \caption{$m_{\chi_1}=11{\rm MeV}$,\,$\delta=1.5{\rm keV}$ at XENONnT {\small (proj.)}}
\end{subfigure}
\begin{subfigure}[b]{0.47\textwidth}
    \includegraphics[width=\textwidth]{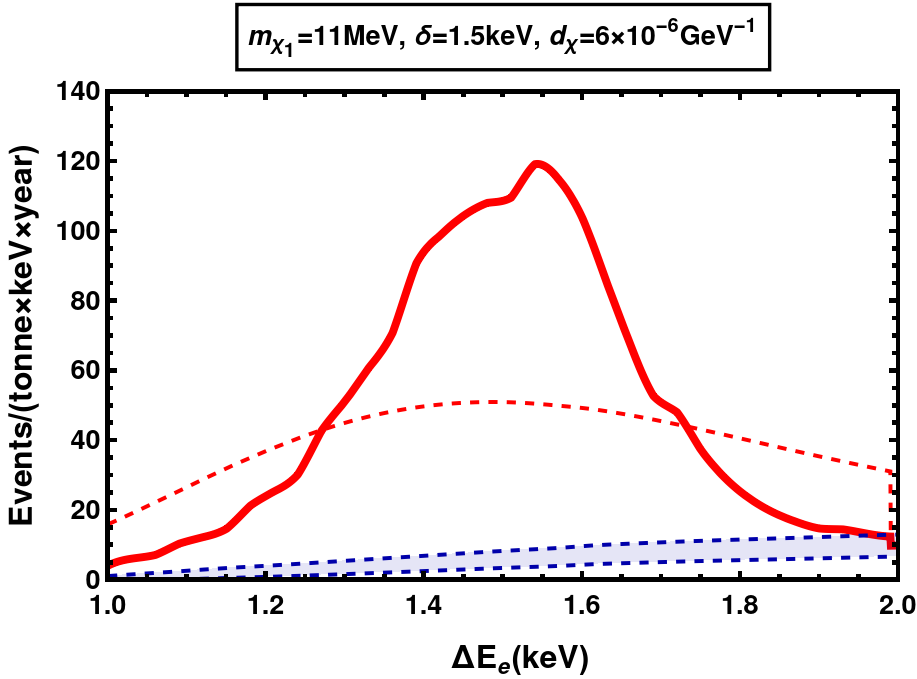}
    \caption{$m_{\chi_1}=11$\,MeV, $\delta=1.5\,$keV at DARWIN}
\end{subfigure}
\caption{Differential event rates for EDM DM. Shown in blue are the background rates with the band representing Poissonian $(\pm\sqrt{N})$ uncertainties. The dashed (solid) red lines show the signal+background rates with (without) smearing from detector resolution, using eq.\,\eqref{eq:e_rec_spec_det} (eq.\,\eqref{eq:e_rec_spec}).}
\label{fig:smeared_spec2}
\end{figure}

We integrate over the theoretical differential event rate as given in eq.\,\eqref{eq:e_rec_spec} to get the total number of events (shown by the color palette in fig.\,\ref{fig:ch4_EDM_res}).  
For comparison with the experimental data, the theoretical spectra can be further smeared using a Gaussian distribution with
energy-dependent width\,\cite{XENON:2020rca,Lee:2020wmh}
\begin{align}
    \frac{dR_D}{\Delta E_e}(\Delta E_e)=\frac{R_D}{\sqrt{2\pi}\sigma(\Delta E_e)}\textrm{exp}\left(-\frac{(\Delta E_e-\delta)^2}{2\sigma(\Delta E_e)^2}\right),
    \label{eq:e_rec_spec_det}
\end{align}
where $\sigma(E)$ is the recoil energy dependent energy resolution of the detector. In figures \ref{fig:smeared_spec1} and \ref{fig:smeared_spec2}, we show the smeared spectra finding that in each case the signal+background rates still shows an excess over the background rates. We use the detector resolution from XENON1T with 
$\sigma(E)=a.\sqrt{E}+b.E$
and $a=0.310\pm0.004\sqrt{\textrm{keV}}$, $b=0.0037\pm0.003$\,\cite{XENON:2020rca}.

\newpage
\bibliographystyle{plain}
\bibliography{Inelastic_FI.bib}

\end{document}